\documentclass[pre,aps,twocolumn]{revtex4} 
\usepackage{epsfig}
\usepackage{amstext}
\usepackage{amsmath}
\usepackage{amsfonts}

\begin{document}

\title{High-Dimensional Inference with the generalized Hopfield Model:\\ 
Principal Component Analysis and Corrections}
\author{S. Cocco $^{1,2}$, R. Monasson $^{1,3}$, V. Sessak $^{3}$}
\affiliation{$^{1}$ Simons Center for Systems Biology, Institute for Advanced Study,  Princeton, NJ 08540, USA\\ $^{2}$Laboratoire de Physique Statistique de l'Ecole Normale Sup\'erieure, CNRS \& Univ. Paris 6, Paris, France\\$^{3}$ Laboratoire de Physique Th\'eorique de l'Ecole Normale Sup\'erieure, CNRS \& Univ. Paris 6,  Paris, France}

\begin{abstract}
We consider the problem of inferring the interactions between a set of $N$ binary variables from the knowledge of their frequencies and pairwise correlations. The inference framework is based on the Hopfield model, a special case of the Ising model where the interaction matrix is defined through a set of patterns in the variable space, and is of rank much smaller than $N$. We show that Maximum Likelihood inference is deeply related to Principal Component Analysis when the amplitude of the pattern components, $\xi$, is negligible compared to $\sqrt N$. Using techniques from statistical mechanics, we calculate the corrections to the patterns to the first order in $\xi/\sqrt N$. We stress that it is important to generalize the Hopfield model and include both attractive and repulsive patterns, to correctly infer networks with sparse and strong interactions. We present a simple geometrical criterion to decide how many attractive and repulsive patterns should be considered as a function of the sampling noise. We moreover discuss how many sampled configurations are required for a good inference, as a function of the system size $N$ and of the amplitude $\xi$. The inference approach is illustrated on synthetic and biological data.
\end{abstract}

\maketitle

\section{Introduction}

Understanding the patterns of correlations between the components of complex systems is a fundamental issue in various scientific fields, ranging from neurobiology to genomic, from finance to sociology, ... A recurrent problem is to distinguish between direct correlations, produced by physiological or functional interactions between the components, and network correlations, which are mediated by other, third-party components. Various approaches have been proposed to infer interactions from correlations, exploiting concepts related to statistical dimensional reduction \cite{jol02}, causality \cite{seth}, the maximum entropy principle \cite{maxent}, Markov random fields \cite{hastie09} ... A major practical and theoretical difficulty in doing so is the paucity and the quality of data: reliable analysis should be able to unveil real patterns of interactions, even if measures are affected by under- or noisy sampling. The size of the interaction network can be comparable to or larger than the number of data, a situation referred to as high-dimensional inference.
 
The purpose of the present work is to establish a quantitative correspondence between two of those approaches, namely the inference of Boltzmann Machines (also called Ising model in statistical physics and undirected graphical models for discrete variables in statistical inference \cite{hastie09}) and Principal Component Analysis (PCA) \cite{jol02}. Inverse Boltzmann Machines (BM) are a mathematically well-founded but computationally challenging approach to infer interactions from correlations. Our scope is to find the interactions among a set of $N$ variables ${\boldsymbol \sigma} =\{\sigma_1,\sigma_2,\ldots , \sigma_N\}$. For simplicity, we consider variables $\sigma_i$ taking binary values $\pm 1$ only; the discussion below can be easily extended to the case of a larger number of values, {\em e.g.} to genomics where nucleotides are encoded by four-letter symbols, or to proteomics where amino-acids can take twenty values. Assume that the average values of the variables, $m_i=\langle \sigma _i\rangle$, and the pairwise correlations, $c_{ij}=\langle\sigma_i\sigma_j\rangle$ are measured, for instance, through the sampling of, say, $B$ configurations ${\boldsymbol \sigma}^b,b=1,\ldots, B$. Solving the inverse BM problem consists in finding the set of interactions, $J_{ij}$, and of local fields, $h_i$, defining an Ising model, such that the equilibrium magnetizations and pairwise correlations coincide with, respectively,  $m_i$ and $c_{ij}$. Many procedures have been designed to tackle this inverse problem, including learning algorithms \cite{ackley}, advanced mean-field techniques \cite{opper,marre}, message-passing procedures \cite{mora,huang}, cluster expansions \cite{noi0,noi}, graphical lasso \cite{hastie09} and its variants \cite{wain}. The performance (accuracy, running time) of those procedures depend on the structure of the underlying interaction network and on the quality of the sampling, {\em i.e.} how large $B$ is. 

Principal Component Analysis (PCA) is a widely popular tool in statistics to analyze the correlation structure of a set of variables ${\boldsymbol \sigma} =\{\sigma_1,\sigma_2,\ldots , \sigma_N\}$. The principle of PCA is simple. One starts with the correlation matrix,
\begin{equation}\label{defgamma}
\Gamma _{ij}=\frac{c_{ij}-m_im_j}{\sqrt{(1-m_i^2)\, (1-m_j^2)}} \ ,
\end{equation}
which expresses the covariance between variables $\sigma_i$ and $\sigma_j$, rescaled by the product of the expected fluctuations of the variables taken separately. $\Gamma$ is then diagonalized. The projections of ${\boldsymbol \sigma}$ along the top eigenmodes (associated to the largest eigenvalues of $\Gamma$) identify the uncorrelated variables which contribute most to the total variance. If a few, say, $p\ (\ll N)$, eigenvalues are notably larger than the remaining ones PCA achieves an important dimensional reduction. The determination of the number $p$ of components to be retained is a delicate issue. It may be done by comparing the spectrum of $\Gamma$ to the Marcenko-Pastur (MP) spectrum for the null hypothesis, that is, for the correlation matrix calculated from the sampling of $B$ configurations of $N$ independent variables \cite{book}. Generally those two spectra coincide when $N$ is large, except for some large or small eigenvalues of $\Gamma$, retained as the relevant components.

The advantages of PCA are multiple, which explains its success. The method is very versatile and fast as it only requires to diagonalize the correlation matrix, which can be achieved in a time polynomial in the size $N$ of the problem. In addition, PCA may be extended to incorporate {\em prior} information about the components, which is particularly helpful for processing noisy data. An illustration is sparse PCA, which looks for principal components with many vanishing entries \cite{spca}. 

In this paper we present a conceptual and practical framework which encompasses BM and PCA in a controlled way. We show that PCA, with appropriate modifications, can be used to infer BM and discuss in detail the amount of data necessary to do so. Our framework is based on an extension of a celebrated model of statistical mechanics, the Hopfield model \cite{hopfield82}. The Hopfield model was originally introduced to model auto-associative memories, and relies on the notion of patterns \cite{amit92}. Informally speaking, a pattern ${\boldsymbol \xi}=(\xi_{1},\ldots , \xi_{N})$ defines an attractive direction in the $N$-dimensional space of the variable configurations, {\em i.e.} a direction along which ${\boldsymbol \sigma}$ has a tendency to align. The norm of ${\boldsymbol \xi}$ characterizes the strength of the attraction. While having only attractive patterns makes sense for auto-associative memories, it is an unnecessary assumption in the context of BM. We therefore generalize the Hopfield model by including repulsive patterns $\hat {\boldsymbol \xi}$, that is, directions in the $N$-dimensional space which ${\boldsymbol \sigma}$ tends to be orthogonal to \cite{nok}. From a technical point of view, the generalized Hopfield model with $p$ attractive patterns and $\hat p$ repulsive patterns is simply a particular case of BM with an interaction matrix, ${\bf J}$, of rank equal to $p+\hat p$. If one knows {\em a priori} that the rank of the true $\bf J$ is indeed small, {\em i.e.} $p+\hat p \ll N$, using the generalized Hopfield model rather than a generic BM allows one to infer much less parameters and to avoid overfitting in the presence of noisy data.

We first consider the case where the components $\xi_{i}$ and $\hat \xi_i$ are very small compared to $\sqrt N$. In this limit case we show that Maximum Likelihood (ML) inference with the generalized Hopfield model is closely related to PCA. The attractive patterns are in one-to-one correspondence with the largest components of the correlation matrix, while the repulsive patterns correspond to the smallest components, which are normally discarded by PCA. When all patterns are selected ($p+\hat p=N$) inference with the generalized Hopfield model is equivalent to the mean-field approximation \cite{opper}. Retaining only few significative components helps, in principle, to remove noise from the data. We present a simple geometrical criterion to decide in practice how many attractive and repulsive patterns should be considered. We also address the question of how many samples ($B$) are required for the inference to be meaningful. We calculate the error bars over the patterns due to the the finite sampling. We then analyze the case where the data are sampled from a generalized Hopfield model, and inference amounts to learn the patterns of that model. When the system size, $N$, and the number of samples, $B$, are both sent to infinity with a fixed ratio, $\alpha=\frac BN$,  there is a critical value of the ratio, $\alpha_c$, below which learning is not possible. The value of $\alpha_c$ depends on the amplitude of the pattern components. This transition corresponds to the retarded learning phenomenon discovered in the context of supervised learning with continuous variables and rigorously studied in random matrix and probability theories, see \cite{engel,book,johnstone06} for reviews. We validate our findings on synthetic data generated from various Ising models with known interactions, and present applications to neurobiological and proteomic data.

In the case of a small system size, $N$, or of very strong components, $\xi _{i},\hat \xi_i$, the ML patterns do not coincide with the components identified by PCA. We make use of techniques from the statistical mechanics of disordered systems originally intended to calculate averages over ensembles of matrices to compute the likelihood to the second order in powers of $\frac {\xi _i}{\sqrt N}$ for a given correlation matrix. We give explicit expressions for the ML patterns in terms of non-linear combinations of the eigenvalues and eigenvectors of the correlation matrix. These corrections are validated on synthetic data. Furthermore, we discuss the issue of how many sampled configurations are necessary to improve over the leading--order ML patterns as a function of the amplitude of the pattern components and of the system size.

The plan of the paper is as follows. In Section \ref{mainres} we define the generalized Hopfield model, the Bayesian inference framework and list our main results, that is, the expressions of the patterns without and with corrections, the criterion to decide the number of patterns, and the expressions for the error bars on the inferred patterns. Tests on synthetic data are presented in Section \ref{secsyn}. Section \ref{secbio} is devoted to the applications to real biological data, {\em i.e} recordings of the neocortical activity of a behaving rat and consensus multi-sequence alignment of the PDZ protein domain family. Readers interested in applying our results rather than in their derivation need not read the subsequent sections. Derivation of the log-likelihood with the generalized Hopfield model and of the main inference formulae can be found in Section \ref{secinf}. In Section \ref{size} we study the minimal number $B$ of samples necessary to achieve an accurate inference, and how this number depends on the number of patterns and on their amplitude. Perspectives and conclusions are given in Section \ref{conc}.

\section{Definitions and main results}
\label{mainres}

\subsection{Generalized Hopfield Model}\label{secgenhop}

We consider configurations $\boldsymbol\sigma=\{\sigma_1,,\sigma_2,\ldots,\sigma_N\}$ of $N$ binary variables taking values $\sigma_i=\pm 1$, drawn according to the probability
\begin{equation}\label{likelihood}
P_H[\boldsymbol\sigma | {\bf h}, \{\boldsymbol \xi^\mu\},\{\hat{\boldsymbol \xi}^\mu\}] = \frac{\exp - E[\boldsymbol\sigma, {\bf h},  \{\boldsymbol \xi^\mu\},\{\hat{\boldsymbol \xi}^\mu\}]}{Z[{\bf h},  \{\boldsymbol \xi^\mu\},\{\hat{\boldsymbol \xi}^\mu\}]} \ ,
\end{equation}
where the energy $E$ is given by
\begin{eqnarray}\label{energy}
E[\boldsymbol\sigma, {\bf h},  \{\boldsymbol \xi^\mu\},\{\hat{\boldsymbol \xi}^\mu\}] &=&-\sum_{i=1}^N h_{i}\sigma _{i} - \frac 1{2N} \sum_{\mu=1}^{p} \left( \sum_{i=1}^N \xi_{i}^\mu\sigma_{i}\right)^2  \nonumber \\
&+& \frac 1{2N} \sum_{\mu=1}^{\hat p} \left( \sum_{i=1}^N \hat\xi_{i}^\mu\sigma_{i}\right)^2 \ .
\end{eqnarray}
The partition function $Z$ in (\ref{likelihood}) ensures the normalization of $P_H$. The components of ${\bf h}=(h_1,h_2,..,h_N)$ are the local fields acting on the variables. The patterns ${\boldsymbol \xi}^\mu=\{\xi_1^\mu,\xi_2^\mu,\ldots,\xi_N^\mu\}$, with $\mu=1,2,\ldots , p$, are attractive patterns: they define preferred directions in the configuration space $\boldsymbol\sigma$, along which the energy $E$ decreases (if the fields are weak enough). The patterns $\hat {\boldsymbol \xi}^\mu$, with $\mu=1,2,\ldots , \hat p$, are repulsive patterns: configurations $\boldsymbol\sigma$ aligned along those directions have a larger energy. The pattern components, $\xi_i^\mu,\hat \xi_i^\mu$, and the fields, $h_{i}$, are real-valued. Our model is a generalized version of the original Hopfield model \cite{hopfield82}, which has only attractive patterns and corresponds to $\hat p=0$. In the following, we will assume that $p+\hat p$ is much smaller than $N$. 

Energy function (\ref{energy}) implicitly defines the coupling $J_{ij}$ between the variables $\sigma_i$ and $\sigma_j$,
\begin{equation}\label{defcoupl}
J_{ij} = \frac 1{N} \sum_{\mu=1}^{p} \xi_{i}^\mu\, \xi_{j}^\mu 
- \frac 1{N} \sum_{\mu=1}^{\hat p} \hat \xi_{i}^\mu\, \hat\xi_{j}^\mu\ .
\end{equation}
Note that any interaction matrix $J_{ij}$ can be written under the form (\ref{defcoupl}), with $p$ and $\hat p$ being, respectively, the number of positive and negative eigenvalues of $J$. Here, we assume that the total number of patterns, $p+\hat p$, {\em i.e.} the rank of the matrix $J$ is (much) smaller than the system size, $N$. 

The data to be analyzed consists of a set of $B$ configurations of the $N$ spins, ${\boldsymbol\sigma}^b$, $b=1,\ldots, B$. We assume that those configurations are drawn, independently from each other, from the distribution $P_H$ (\ref{likelihood}). The parameters defining $P_H$, that is, the fields ${\bf h}$ and the patterns $\{{\boldsymbol\xi}^\mu\},\{\hat{\boldsymbol\xi}^\mu\}$ are unknown. Our scope is to determine the most likely values for those fields and patterns from the data. In Bayes inference framework the posterior distribution for the fields and the patterns given the data $\{{\boldsymbol\sigma}^b\}$ is
\begin{eqnarray}\label{post}
P[{\bf h}, \{\boldsymbol \xi^\mu\},\{\hat{\boldsymbol \xi}^\mu\}| \{\boldsymbol \sigma ^b\}]&=&  \frac{P_{0}[{\bf h}, \{\boldsymbol \xi^\mu\},\{\hat{\boldsymbol \xi}^\mu\}]}{P_{1}[\{\boldsymbol \sigma ^b\}]}  \\
&\times&\prod _{b=1}^B P_H[\boldsymbol\sigma ^b | {\bf h}, \{\boldsymbol \xi^\mu\},\{\hat{\boldsymbol \xi}^\mu\}]\ , \nonumber
\end{eqnarray}
where $P_0$ encodes some {\em a priori} information over the parameters to be inferred and $P_1$ is a normalization. 

It is important to realize that many transformations affecting the patterns can actually leave the coupling matrix ${\bf J}$ (\ref{defcoupl}) and the distribution $P_H$ unchanged. A simple example is given by an orthogonal transformation ${\cal O}$ over the attractive patterns : $\xi_i^{\mu}\to \bar \xi_i^{\mu} = \sum_\nu {\cal O}^{\mu\nu} \xi_i^\nu$. This invariance entails that the the problem of inferring the patterns is not statistically consistent: even with an infinite number of sampled data no inference procedure can distinguish between a Hopfield model with patterns $\{\boldsymbol\xi ^\mu \}$ and another one with patterns $\{ \bar{\boldsymbol \xi} ^\mu\}$. However, the inference of the couplings is statistically consistent: two distinct matrices ${\bf J}$ define two distinct distributions over the data. 

In the presence of repulsive patterns the complete invariance group is the indefinite orthogonal group $O(p,\hat p)$, which has $\frac 12 (p+\hat p)(p+\hat p-1)$ generators. To select one particular set of most likely patterns, we explicitly break the invariance through $P_0$. A convenient choice we use throughout this paper is to impose that the weighted dot products of the pairs of attractive and/or repulsive patterns vanish:
\begin{eqnarray}\label{gauge}
\sum_i \xi_i^\mu \xi_i^\nu(1-m_i^2) &=& 0 \quad  \bigg[\frac 12 p(p-1)\ \hbox{\rm constraints}\bigg]\ , \nonumber \\
\sum_i \xi_i^\mu \hat \xi_i^\nu(1-m_i^2)  &=& 0 \quad \bigg[ p \hat p\ \hbox{\rm constraints}\bigg]\ , \\
\sum_i \hat \xi_i^\mu \hat \xi_i^\nu(1-m_i^2)&=& 0 \quad \bigg[\frac 12 \hat p(\hat p-1)\ \hbox{\rm constraints}\bigg]\ . \nonumber 
\end{eqnarray}
In the following we will use the vocable Maximum Likelihood inference to refer to the case where the prior $P_{0}$ is used to break the invariance only. $P_0$ may also be chosen to impose specific constraints on the pattern amplitude, see Section \ref{secregu} devoted to regularization.

\subsection{Maximum Likelihood Inference: lowest order}\label{sec0}

Due to the absence of three- or higher order-body interactions in $E$ (\ref{energy}), $P$ depends on the data $\{{\boldsymbol\sigma}^b\}$ only through the $N$ magnetizations, $m_i$, and the $\frac 12N(N-1)$ two-spin covariances, $c_{ij}$, of the sampled data: 
\begin{equation}
m_i = \frac 1B \sum_{b} \sigma_i^b \quad , \qquad 
c_{ij} = \frac 1B \sum_{b} \sigma_i^b\, \sigma_{j}^b \ .
\end{equation}
We consider the correlation matrix $\Gamma$ (\ref{defgamma}), and call $\lambda^1\ge \ldots\ge \lambda^k\ge \lambda^{k+1} \ge \ldots\ge \lambda^N$ its eigenvalues. ${\bf v}^k$ denotes the eigenvector attached to $\lambda^k$  and normalized to unity. We also introduce another notation to label the same eigenvalues and eigenvectors in the reverse order: $\hat \lambda^k \equiv \lambda^{N+1-k}$ and $\hat {\bf v}^k = {\bf v}^{N+1-k}$, {\em e.g.} $\hat \lambda^1$ is the smallest eigenvalue of $\Gamma$; the motivation for doing so will be transparent below. 
Note that $\Gamma$ is, by construction, a semi-definite positive matrix: all its eigenvalues are positive. In addition, the sum of the eigenvalues is equal to $N$ since $\Gamma_{ii}=1, \forall i$. Hence the largest and smallest eigenvalues are guaranteed to be, respectively, larger and smaller than unity. 

In the following Greek indices, {\em i.e.} $\mu,\nu,\rho$, correspond to integers comprised between 1 and $p$ or $\hat p$, while roman letters, {\em i.e.} $i,j,k$ denote integers ranging from 1 to $N$.

Finding the patterns and fields maximizing $P$ (\ref{post}) is a very hard computational task. We introduce an approximation scheme for those parameters
\begin{eqnarray}\label{correxp}
\xi_i^\mu &=& (\xi^0)^\mu_i +  (\xi^1)^\mu_i + \ldots \ , \nonumber \\
\hat \xi_i^\mu &=& (\hat \xi^0)^\mu_i +  (\hat \xi^1)^\mu_i + \ldots \ , \nonumber\\
h_i&=& (h^0)_i +  (h^1)_i + \ldots \ .
\end{eqnarray}
The derivation of this systematic approximation scheme and the discussion of how smaller the contributions get with the order of the approximation can be found in Section \ref{exp67}. To the lowest order the patterns are given by
\begin{eqnarray} \label{ordre0xi}
(\xi ^0)_i^\mu &=& \sqrt{N\, \left( 1 - \frac 1{\lambda^\mu}\right)}\;\frac {v_i^\mu}{\sqrt{1-m_i^2}} \quad (1\le\mu\le p)\\
(\hat \xi ^0)_i^\mu &=& \sqrt{N\, \left( \frac 1{\hat \lambda^{\mu}}-1\right)}\;\frac {\hat v_i^{\mu}}{\sqrt{1-m_i^2}} \quad (1\le\mu\le\hat p)\nonumber 
\ .
\end{eqnarray}
The above expressions require that $\lambda^\mu>1$ for an attractive pattern and $\hat\lambda^\mu<1$ for a repulsive pattern. Once the patterns are computed the interactions, $(J^0)_{ij}$, can be calculated from (\ref{defcoupl}),
\begin{eqnarray}\label{j0}
(J^0)_{ij} &=& \frac {1}{\sqrt{(1-m_i^2)(1-m_j^2)}} \left(\sum_{\mu=1}^{p} \left(1 -\frac 1{\lambda^\mu}\right) v_{i}^\mu\, v_{j}^\mu \right. \nonumber \\
&-& \left. \sum_{\mu=1}^{\hat p} \left(\frac 1{\hat \lambda^\mu}-1\right) \hat v_{i}^\mu\, \hat v_{j}^\mu\right)\ .
\end{eqnarray}
The values of the local fields are then obtained from 
\begin{equation}\label{ordre0hbis}
(h^0)_i = \tanh ^{-1} m_i - \sum _{j} (J^0)_{ij} \,m_j \ ,
\end{equation}
which has a straightforward mean-field interpretation.
 
The above results are reminiscent of PCA, but differ in several significative aspects. First, the patterns do not coincide with the eigenvectors due to the presence of $m_i$-dependent terms. Secondly, the presence of the $\lambda^\mu$-dependent factor in (\ref{ordre0xi}) discounts the patterns corresponding to eigenvalues close to unity. This effect is easy to understand in the limit case of independent spins and perfect sampling ($B\to\infty$): $\Gamma$ is the identity matrix, which gives $\lambda^\mu=1, \forall \mu$, and the patterns rightly vanish. Thirdly, and most importantly, not only the largest but also the smallest eigenmodes must be taken into account to calculate the interactions. 

The couplings $J^0$ (\ref{j0}) calculated from the lowest-order approximation for the patterns are closely related to the mean-field (MF) interactions \cite{opper},
\begin{equation}\label{jtap}
J^{MF}_{ij} = - \frac { (\Gamma^{-1})_{ij}}{\sqrt{(1-m_i^2)(1-m_j^2)}} \ ,
\end{equation}
where $\Gamma^{-1}$ denotes the inverse matrix of $\Gamma$ (\ref{defgamma}). However, while all the eigenmodes of $\Gamma$ are taken into account in the MF interactions (\ref{jtap}), our lowest-order interactions (\ref{j0}) include contributions from the $p$ largest and the $\hat p$ smallest eigenmodes only. As the values of $p,\hat p$ can be chosen depending on the number of available data, the generalized Hopfield interactions (\ref{j0}) is {\em a priori} less sensitive to overfitting. In particular, it is possible to avoid considering the bulk part of the spectrum of $\Gamma$, which is essentially due to undersampling (\cite{book} and Section \ref{rl}).

\subsection{Sampling error bars on the patterns}\label{secsam}

The posterior distribution $P$ can locally be approximated with a Gaussian distribution centered in the most likely values for the patterns, $\{(\boldsymbol\xi^0)^\mu\}, \{(\hat{\boldsymbol\xi} ^0)^\mu\}$, and the fields, ${\bf h}^0$. We obtain the covariance matrix of the fluctuations of the patterns around their most likely values,
\begin{equation}
\langle \Delta \xi_i^\mu  \,\Delta \xi_j^\nu\rangle = \frac {N\, \big[{\bf M}_{\xi\xi}\big]_{ij}^{\mu \nu}}{B\sqrt{(1-m_i^2)(1-m_j^2)}}\; \ .
\label{errorbar}
\end{equation}
and identical expressions for $\langle \Delta \xi_i^\mu  \,\Delta \hat \xi_j^\nu\rangle$ and $\langle \Delta \hat \xi_i^\mu  \,\Delta \hat \xi_j^\nu\rangle$ upon substitution of $\big[ {\bf M}_{\xi\xi}\big]_{ij}^{\mu \nu}$ with, respectively, $\big[{\bf M}_{\xi\hat\xi}\big]_{ij}^{\mu \nu}$ and $\big[{\bf M}_{\hat\xi\hat\xi}\big]_{ij}^{\mu \nu}$. The entries of the $\bf M$ matrices are 
\begin{eqnarray}\label{errorbar2}
 \big[ {\bf M}_{\xi\xi}\big]_{ij}^{\mu \nu}\!\!\!&=&\!\! \delta^{\mu\nu}\!\! \left[ \sum _{k=p+1}^{N-\hat p} \frac{v_i^k\,v_j^k}{|\lambda^k-\hat\lambda^\mu|} + \sum _{\rho =1} ^{p} \frac{|\lambda^\mu-1| \lambda^\rho\, v_i^\rho\, v_j^\rho }{G_1(\lambda ^\rho,\lambda^\mu)}\right.\nonumber\\
&+&\left. \sum _{\rho=1}^{\hat p} \frac{|\lambda^\mu-1|\hat\lambda^\rho\, \hat v_i^\rho\, \hat v_j^\rho }{G_1(\hat \lambda ^\rho,\lambda^\mu)}\right]+ \frac{G_2( \lambda^\mu, \lambda^\nu)}{G_1( \lambda^\mu, \lambda^\nu)}\,  v_j^\mu\, v_i^\nu \ ,\nonumber \\
 \big[ {\bf M}_{\xi\hat\xi}\big]_{ij}^{\mu \nu}\!\!\!&=&\!\! \frac{G_2(\lambda^\mu,\hat \lambda^\nu)}{G_1(\lambda^\mu,\hat \lambda^\nu)} \, v_j^\mu\, \hat v_i^\nu\ , 
\end{eqnarray}
and $\big[ {\bf M}_{\hat\xi\hat\xi}\big]_{ij}^{\mu \nu}$ is obtained from $\big[{\bf M}_{\xi\xi}\big]_{ij}^{\mu \nu}$ upon substitution of $\lambda^\mu,\lambda^\nu,v_i^\mu,v_i^\nu$ with, respectively, $\hat \lambda^\mu,\hat \lambda^\nu,\hat v_i^\mu,\hat v_i^\nu$. 
Functions $G_1$ and $G_2$ are defined through
\begin{eqnarray}
G_1(x,y) &=& (x\,|y-1| +y\,|x-1|)^2 \ ,\nonumber \\
G_2(x,y) &=& \sqrt{x\, y \, |x-1|\, |y-1|} \ .
\end{eqnarray}
The covariance matrix of the fluctuations of the fields is given in Section \ref{secerrcal}. Error bars on the couplings (\ref{defcoupl}) can be calculated from the ones on the patterns.

Formula (\ref{errorbar}) tells us how significative are the inferred values of the patterns in the presence of finite sampling. For instance, if the error bar $\langle (\Delta \xi_i^\mu)^2\rangle ^{1/2}$ is larger than, or comparable with the pattern component $(\xi^0)_i^\mu$ calculated from (\ref{ordre0xi}) then this component is statistically compatible with zero. According to formula (\ref{errorbar}) we expect error bars of the order of $\frac 1{\sqrt \alpha}$ over the pattern components, where $\alpha =\frac BN$. 

\subsection{Optimal numbers of attractive and repulsive patterns}
\label{secresupatt}

\begin{figure}
\begin{center}
\epsfig{file=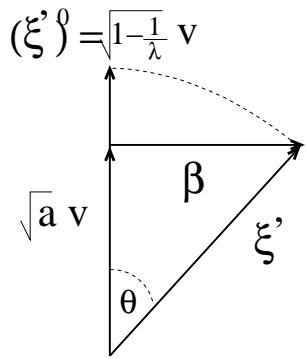,width=4.cm}
\caption{Geometrical representation of identity (\ref{fluc78}), expressing the rescaled pattern ${\boldsymbol \xi}'$ as a linear combination of the eigenvector ${\bf v}$ and of the orthogonal fluctuations ${\boldsymbol \beta}$. The most likely rescaled pattern, $({\boldsymbol\xi} ')^0$, corresponds to $a=1-\frac 1\lambda, \boldsymbol\beta=0$.The dashed arc has radius $\sqrt{1-\frac1{\lambda}}$. The subscript $\mu$ has been dropped to lighten notations.}
\label{fig-schema0}
\end{center}
\end{figure}

We now determine the numbers of patterns, $p$ and $\hat p$, based on a simple geometric criterion; the reader is referred to Section \ref{secoptim} for detailed calculations. To each attractive pattern $\boldsymbol\xi^\mu$ we associate the rescaled pattern $(\boldsymbol \xi ^\mu)'$, whose components are $(\xi_i^\mu)'=\xi_i^\mu \sqrt{1-m_i^2}/\sqrt{N}$. We write 
\begin{equation}\label{fluc78}
(\boldsymbol\xi ^\mu) ' = \sqrt {a^\mu} \; {\bf v}^\mu + \boldsymbol\beta^\mu \ ,
\end{equation}
where $a^\mu$ is a positive coefficient, and $\boldsymbol\beta^\mu $ is a vector orthogonal to all rescaled patterns by virtue of (\ref{gauge}) (Fig.~\ref{fig-schema0}). Our lowest order formula (\ref{ordre0xi}) for the Maximum Likelihood estimators gives $a^\mu=1-\frac 1{\lambda^\mu }$ and $\boldsymbol\beta^\mu  =0$, see Fig.~\ref{fig-schema0}. This result is, to some extent, misleading. While the most likely value for the vector $\boldsymbol \beta^\mu$ is indeed zero, its norm is almost surely not vanishing! The statement may appear paradoxical but is well-known to hold for stochastic variables: while the average or typical value of the location of an isotropic random walk vanishes, its average squared displacement does not. Here, $ {\boldsymbol \beta}^\mu $ represents the stochastic difference between the pattern to be inferred and the direction of one of the largest eigenvectors of $\Gamma$. We expect the squared norm $(\boldsymbol\beta^\mu )^2$ to have a non-zero value in the $N,B\to\infty$ limit at fixed ratio $\alpha=\frac BN >0$. Its average value can be straightforwardly computed from formula (\ref{errorbar2}),
\begin{equation}\label{amumap2c0}
\langle (\boldsymbol\beta^\mu )^2 \rangle = \frac 1B\; \sum_i \big[ M_{\xi\xi}\big]_{ii}^{\mu \mu} =\frac 1B\sum _{k=p+1}^{N-\hat p} \frac 1{\lambda^\mu-\lambda^k} \ ,  
\end{equation}
where $\mu$ is the index of the pattern. We define the angle $\theta^\mu$ between the eigenvector ${\bf v}^\mu$ and the rescaled pattern $({\boldsymbol \xi}^\mu)'$ through
\begin{equation}
\theta^\mu = \sin^{-1} \sqrt{ \frac{\langle (\boldsymbol\beta^\mu)^2\rangle}{1-\frac 1{\lambda^\mu}}} \ ,
\label{gc}
\end{equation}
see Fig.~\ref{fig-schema0}. Small values of $\theta^\mu$ correspond to reliable patterns, while large $\theta^\mu$ indicate that the Maximum Likelihood estimator of the $\mu^{th}$ pattern is plagued by noise. The value of $p$ such that $\theta^p$ is, say, about $\frac \pi 4$ is our estimate for the number of attractive patterns. 

The above approach can be easily repeated in the case of repulsive patterns. We obtain, with obvious notations,
\begin{equation}\label{amumap2cr}
\langle (\hat{\boldsymbol\beta}^\mu )^2 \rangle = \frac 1B\; \sum_i \big[ M_{\hat\xi\hat\xi}\big]_{ii}^{\mu \mu} =\frac 1B\sum _{k=p+1}^{N-\hat p} \frac 1{\lambda^k-\hat\lambda^\mu} \ ,  
\end{equation}
and
\begin{equation}
\hat \theta^\mu = \sin^{-1} \sqrt{ \frac{\langle (\hat{\boldsymbol\beta}^\mu)^2\rangle}{\frac 1{\hat\lambda^\mu}-1}} \ .
\label{gcr}
\end{equation}
The value of $\hat p$ such that $\hat \theta^{\hat p}$ is, say, about $\frac \pi 4$ is our estimate for the number of repulsive patterns. 

\subsection{Regularization}\label{secregu}

So far we have considered that the prior probability $P_0$ over the patterns was uniform, and was used to break the invariance through the conditions (\ref{gauge}). The prior probability can be used to constrain the amplitude of the patterns. For instance, we can introduce a Gaussian prior on the patterns,
\begin{equation}\label{regu1}
P_0 \propto\exp \left[ -\frac {\gamma}2\sum_{i=1}^N  (1-m_i^2) \left(\sum_{\mu=1}^p (\xi_i^\mu)^2+ \sum_{\mu=1}^{\hat p} (\hat \xi_i^\mu)^2 \right)\right] ,
\end{equation}
which penalizes large pattern components \cite{noi}. The presence of the $(1-m_i^2)$ factor entails that the effective strength of the regularization term, $\gamma (1-m_i^2)$, depends on the site magnetization. Regularization is particularly useful in the case of severe undersampling. With regularization (\ref{regu1}) the lowest order expression for the pattern is still given by (\ref{ordre0xi}), after carrying out the following transformation on the eigenvalues,
\begin{eqnarray}\label{transformeigen}
\lambda ^\mu &\to& \lambda^\mu -\gamma \ , \quad (\mu =1,\ldots, p) \ ,\nonumber \\
\lambda ^k &\to& \lambda^k \ , \quad\qquad (k =p+1,\ldots, N-\hat p) \ ,\nonumber \\
\hat \lambda ^\mu &\to& \hat \lambda^\mu +\gamma \ , \quad (\mu =1,\ldots, \hat p) \ .
\end{eqnarray}
The values of $p$ and $\hat p$ must be such that the transformed $\lambda^p$ and $\hat \lambda ^{\hat p}$ are, respectively, larger and smaller than unity. Regularization (\ref{regu1}) ensures that the couplings do not blow up, even in the presence of zero eigenvalues in $\Gamma$. Applications will be presented in Sections \ref{secsyn} and \ref{secbio}. The value of the regularization strength $\gamma$ can be chosen based on a Bayesian criterion \cite{McKay}.

\subsection{Maximum likelihood inference: first corrections}\label{sec1}

We now give the expression for the first-order correction to the attractive patterns,
\begin{equation}\label{ordre1xi}
(\xi^1)_i^\mu =\sqrt{\frac N{1-m_i^2}} \sum_{k=1}^N  A^{k \mu}\; B^{k\mu}\; v_{i}^k\ ,
\end{equation}
where
\begin{eqnarray}\label{defB}
A^{k \mu}&=& C^k C^\mu + \left( \sum_{\rho=1}^p +  \sum_{\rho=N+1-\hat p}^N \right) (\lambda^\rho-1)\nonumber \\ &\times& \sum_{i} v_{i}^k v_{i}^\mu
\bigg[ (v_i^\rho)^2 + \frac{2\,m_i\,C^\rho\, v_i^\rho}{\sqrt{1-m_i^2}}\,  \bigg] 
\end{eqnarray}
and 
\begin{equation}\label{defC}
B^{k \mu} = \left\{ \begin{array}{c c c}
\frac 12 \sqrt{\frac{\lambda^\mu}{\lambda^\mu-1}}  & \hbox{\rm if}& k \le p \ ,\\ & & \\
\frac{\sqrt{\lambda^\mu(\lambda^\mu-1)}}{\lambda^\mu -\lambda ^k} & \hbox{\rm if}
&k \ge p+1 \  .
\end{array}\right. 
\end{equation}
and
\begin{equation}\label{defC3}
C^k = \sum _i \frac{m_i\, v_i^k}{\sqrt{1-m_i^2}} \left( \sum_{\rho=1}^p +  \sum_{\rho=N+1-\hat p}^N \right)(\lambda^\rho-1) \, (v_i^\rho)^2\ .
\end{equation}
Similarly, the first corrections to the repulsive patterns are
\begin{equation}\label{ordre1xir}
(\hat\xi^1)_i^\mu =\sqrt{\frac N{1-m_i^2}} \sum_{k=1}^N  \hat A^{k \mu}\; \hat B^{k\mu}\; v_{i}^k\ .
\end{equation}
The definition of $\hat A^{k\mu}$ is identical to (\ref{defB}), with $C^\mu$ and $v_i^\mu$ replaced with, respectively, $C^{N+1-\mu}$ and $\hat v_i^{\mu}$. Finally,
\begin{equation}\label{defCr}
\hat B^{k \mu} = \left\{ \begin{array}{c c c}
\frac 12 \sqrt{\frac{\hat \lambda^\mu}{1-\hat\lambda^\mu}}  & \hbox{\rm if}& k \ge N-\hat p+1 \ ,\\ & & \\
\frac{\sqrt{\hat \lambda^\mu(1-\hat\lambda^\mu)}}{\hat\lambda ^\mu - \lambda^k} & \hbox{\rm if}
&k \le N-\hat p \  .
\end{array}\right. 
\end{equation}
The first order corrections to the fields $h_i$ can be found in Section \ref{foh}.

It is interesting to note that the corrections to the pattern ${\boldsymbol \xi}^\mu$ involve non-linear interactions between the eigenmodes of $\Gamma$. Formula (\ref{defB}) for $A^{k\mu}$ shows that the modes $\mu$ and $k$ interact through a multi-body overlap with mode $\rho$ (provided $\lambda^\rho \ne 1$). In addition, $A^{k\mu}$ does not {\em a priori} vanish for $k\ge p+1$: corrections to the patterns have non--zero projections over the 'noisy' modes of $\Gamma$. In other words, valuable information over the true values of the patterns can be extracted from the eigenmodes of $\Gamma$ associated to bulk eigenvalues. 

\subsection{Quality of the inference vs. size of the data set}\label{secqual}

\begin{figure}[b]
\begin{center}
\epsfig{file=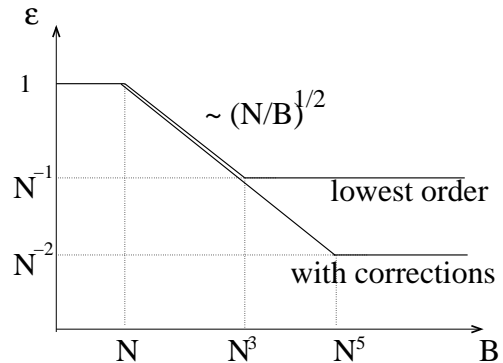,width=6.5cm}
\caption{Schematic behaviour of the error $\epsilon$ on the inferred patterns as a function of the number $B$ of sampled configurations and for a problem size equal to $N$, when the pattern components are of the order of unity compared to $N$. See main text for the case of few large pattern components, of the order of $\sqrt N$, {\em i.e.} couplings $J$ of the order of 1.}
\label{fig-symbolsum}
\end{center}
\end{figure}

The accuracy $\epsilon$ on the inferred pattern is limited both by the sampling error resulting from the finite number of data and the intrinsic error due to the expansion (\ref{correxp}). According to Section \ref{secsam}, the sampling error on the pattern component is expected to decrease as $\sim \sqrt{\frac NB}$. The intrinsic error depends on the order of the expansion, on the size $N$ and on the amplitude of the patterns.

No inference is possible unless the ratio $\alpha = \frac BN$ exceeds a critical value, referred to as $\alpha_c$ in the following (Section \ref{weak1}). This phenomenon is similar to the retarded learning phase transition discovered in the context of unsupervised learning \cite{engel}.

Assume that the pattern components $\xi_{i}$ are of the order of one (compared to $N$), that is, that the couplings are almost all non zero and of the order of $\frac 1N$. Then, the intrinsic error is of the order of $\frac 1N$ with the lowest order formula (\ref{ordre0xi}), and of the order of $\frac 1{N^2}$ when corrections (\ref{ordre1xi}) are taken into account; for a more precise statement see Section \ref{exp67} and formula (\ref{expr}). The corresponding values of $B$ at which saturation takes place are, respectively, of the order of $N^{3}$ and $N^5$. The behaviour of the relative error between the true and inferred patterns, $\epsilon$ (\ref{epsJ}), is summarized in Fig.~\ref{fig-symbolsum}. In general we expect that $B\sim N^{1+2a}$ samples at least are required to have a more accurate inference with $a^{th}$-order patterns than with $(a-1)^{th}$-order patterns. Furthermore there is no need to sample more than $N^{3+2a}$ configurations when using the $a^{th}$-order expression for the patterns. 

If the system has $O(N)$ non vanishing couplings $J_{ij}$ of the order of $J$, then patterns have few large components, of the order of $\sqrt J$. In this case the intrisic error over the patterns will be of the order of $J$ with the lowest order inference formulae, and of the order of $J^2$ with the first corrections. The numbers of sampled configurations, $B$, required to reach those minimal errors will be, respectively, of the order of $\frac N{J^2}$ and $\frac N{J^4}$.

\section{Tests on synthetic data}\label{secsyn}

In this Section we test the formulae of Section \ref{mainres} for the patterns and fields against synthetic data generated from various Ising models with known interactions. We consider four models:
\begin{itemize}
\item{\em Model A} is a Hopfield model with $N=100$ spins, $p$ (= 1 or 3) attractive patterns and no repulsive pattern ($\hat p=0$). The components of the patterns are Gaussian random variables with zero mean and standard deviation $\xi$, specified later. The local fields $h_i$ are set equal to zero. 
\item{\em Model B:} Model B consists of $N$ spins, grouped into four blocks of $\frac N4$ spins each. The $p=3$ patterns have uniform components over the blocks: $\xi^1=\frac {2\sqrt 3}5(0,1,1,1)$, $\xi^2 =\frac 25 (\sqrt 3,1,-2,1)$, $\xi^3 = \frac 25(\sqrt 3,-2,1,1)$. The fields are set to zero. Those choices ensure that the pattern are orthogonal to each other, and have a weak intensity: on average, $|\xi|^2=\frac 9{25} < 1$. 
\item{\em Model C} is a very simple Ising model where all fields and couplings vanish, except coupling $J_{12}\equiv J$ between the first two spins. 
\item{\em Model D} is an Ising model with $N=50$ spins, on an Erdos-Renyi random graph with average connectivity (number of neighbors for each spin) equal to $d=5$ and coupling values $J$ distributed uniformly between -1 and 1. Model D is an instance of the Viana-Bray model \cite{vb}. In the thermodynamic limit $N\to\infty$ this model is in the spin glass phase since $d \langle \tanh^2(J)\rangle_J >1$ \cite{vb}.
\end{itemize}

For each one of the models above, the magnetizations and pairwise correlations can be estimated through the sampling of $B$ configurations at equilibrium using Monte Carlo simulations. This allows us to estimate the consequence of sampling noise on the inference quality by varying the value of $B$. Furthermore, for models $B$ and $C$, it is possible to obtained the exact Gibbs values for $m_i$ and $c_{ij}$ (corresponding to a perfect sampling, $B=\infty$)\footnote{As a result of the block structure the energy (\ref{energy}) depends on the $N$--spin configuration through the four block magnetizations (sums of the $\frac N4$ spins in each block) only. Hence, the correlations $c_{ij}$ and magnetizations $m_i$ can be calculated in a time growing as $N^4$ (instead of $2^N$), which allows us to reach sizes equal to a few hundreds easily.}. This allows us to study the systematic error resulting from formulae (\ref{ordre0xi},\ref{ordre1xi},\ref{ordre1xir}), irrespectively of the sampling noise. 

Model A is used to test the lower order formula for the patterns, and how the quality of inference depends on the amplitude of the patterns. Models C and D are highly diluted networks with strong $J=O(1)$ interactions, while models A and B correspond to dense networks with weak $J=O(\frac 1N)$ couplings. Models C and D are, therefore, harder benchmarks for the generalized Hopfield model. In addition, the couplings implicitly define, through (\ref{defcoupl}), both attractive and repulsive patterns. Those models can thus be used to determine how much repulsive patterns are required for an accurate inference of general Ising models. 

\subsection{Dominant order formula for the patterns}

\begin{figure}
\begin{center}
\epsfig{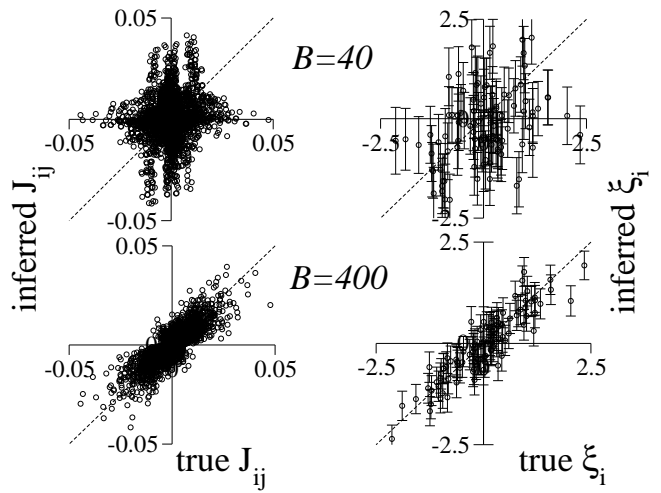}
\caption{Application of formula (\ref{ordre0xi}) to two sets of $B=40$ (top) and 400 (bottom) configurations, randomly generated from the distribution $P_H$ (\ref{likelihood}) for model A with $p=1$ pattern. The standard deviation of the pattern components is $\xi=.7$. {\bf Left:} comparison of the true and inferred couplings for each pair $(i,j)$. {\bf Right:} comparison of the true and inferred components $\xi_i$ of the pattern, with the error bars calculated from (\ref{errorbar}). The dashed lines have slope unity. Inference is done with $p=1$ attractive pattern and no repulsive pattern.}
\label{fig-compar-1patt}
\end{center}
\end{figure}

\begin{figure}
\begin{center}
\epsfig{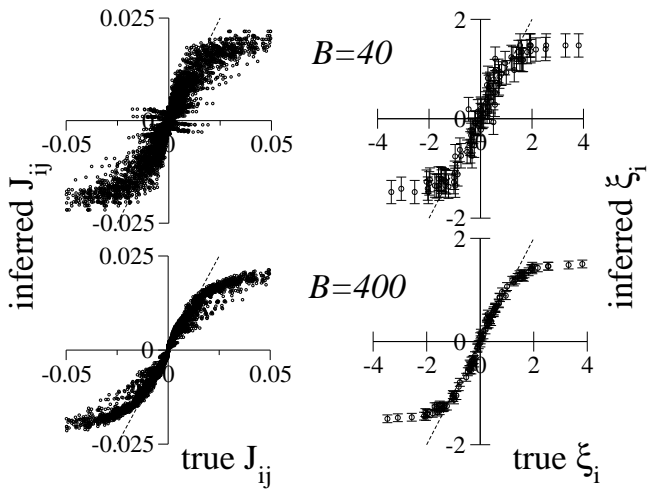}
\caption{Same as Fig.~\ref{fig-compar-1patt}, but with a standard deviation $\xi=1.3$ instead of $\xi=.7$. The amplitude is strong enough to magnetize the configurations along the pattern, see Sections \ref{strong1} and \ref{secferro2}.}
\label{fig-compar-1patt-bis}
\end{center}
\end{figure}

We start with Model A with $p=1$ pattern. In this case, no ambiguity over the inferred pattern is possible since the energy $E$ is not invariant under continuous transformations, see Section \ref{secgenhop}. We may therefore directly compare the true and the inferred patterns. Figures~\ref{fig-compar-1patt} and \ref{fig-compar-1patt-bis} show the accuracy of the lowest order formula for the patterns, eqn (\ref{ordre0xi}). If the pattern components are weak, each sampled configuration $\boldsymbol \sigma$ is weakly aligned along the pattern $\boldsymbol\xi$. If the number $B$ of sampled configurations is small, the largest eigenvector of $\Gamma$ is uncorrelated with the pattern direction (Fig.~\ref{fig-compar-1patt}). When the size of the data set is sufficiently large, {\em i.e.} $B>\alpha_{c} N$ (Section \ref{weak1}), formula (\ref{ordre0xi}) captures the right direction of the pattern, and the inferred couplings are representative of the true interactions. Conversely, if the amplitudes of the components of the pattern $\boldsymbol\xi$ are strong enough, each sampled configuration $\boldsymbol\sigma$ is likely to be aligned along the pattern. A small number $B$ (compared to $N$) of those configurations suffice to determine the pattern (Fig.~\ref{fig-compar-1patt-bis}). In the latter case, we see that the largest components $\xi_i$ are systematically underestimated. A systematic study of how large $B$ should be for the inference to be reliable can be found in Section \ref{size}.

\begin{figure}
\begin{center}
\epsfig{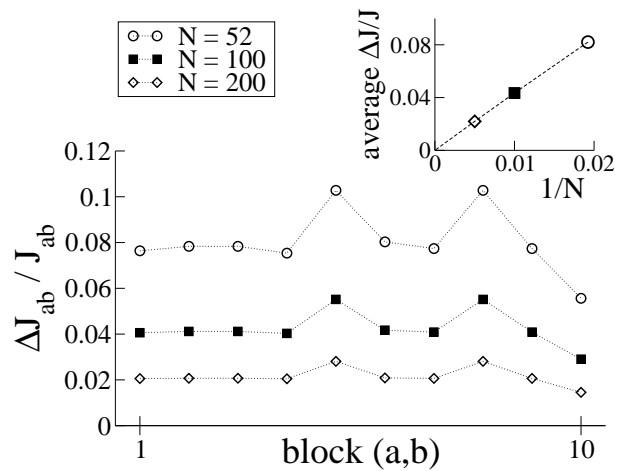}
\caption{Relative differences between the true and the inferred couplings, $\Delta J_{ab}/J_{ab}$ for three system sizes, $N$. The inference was done using the lowest order ML formulae (\ref{ordre0xi}) for the patterns. Data were generated from Model B  (perfect sampling); there are {\em a priori} ten distinct values of the couplings, one for each pair of blocks $a$ and $b$.  Inset: average value of $\Delta J_{ab}/J_{ab}$ as a function of $\frac 1N$. Circles, squares and diamonds correspond to, respectively, $N=52$, 100 and 200 spins.}
\label{fig-compar-3patt}
\end{center}
\end{figure}

We now use model B to generate the data. As model $B$ includes more than one pattern, the inferred patterns cannot be compared to the true one easily due to the invariance of Section \ref{secgenhop}. We therefore compare in Fig.~\ref{fig-compar-3patt} the true couplings and the interactions found using (\ref{ordre0xi}) for three sizes, $N=52$, $100$ and $200$. The size $N$ sets also the amplitude of the couplings, which decreases as $\frac 1N$ from (\ref{defcoupl}). As the patterns are uniform among each one of the four blocks there are ten possibles values for the couplings $J_{ij}$, depending on the labels $a$ and $b$ of the blocks to which $i$ and $j$ belong, with $1\le a \le b\le 4$. For $N=100$ spins, the relative errors range between 3 and 5.5\%. When the number of spins is doubled (respectively, halved) the relative errors are about twice smaller (respectively, larger). This result confirms that formula (\ref{ordre0xi}) is exact in the infinite $N$ limit only, and that corrections of the order of $O(\frac 1N)$ are expected for finite system sizes (Inset of Fig.~\ref{fig-compar-3patt}). This scaling was expected from Section \ref{secqual}.

\begin{figure}
\begin{center}
\vskip 1cm
\epsfig{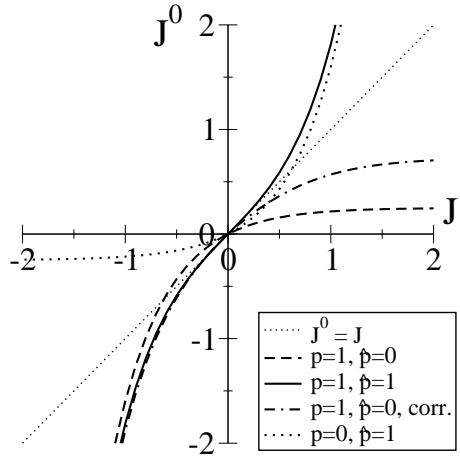}
\vskip .5cm
\caption{Inferred coupling $J^0$ between the first two spins of Model C, within lowest order ML, and as a function of the true coupling $J$. Values of $p$ and $\hat p$ are shown in the Figure.}
\label{infj12}
\end{center}
\end{figure}

We now consider model C. For perfect sampling ($B=\infty$) the correlation matrix (\ref{defgamma}) is
\begin{equation}
\Gamma= \left( \begin{array}{ccccc}
1 &\mbox{ $\tanh J$} &0& \ldots &0 \\\mbox{$\tanh J$} & 1 &0& \ldots & 0 \\
0  &0&1& \ldots & 0 \\ 0 &\ldots& 0& 1 & 0 \\0 & $\ldots$ & 0&0 &1 \end{array} \right)\ .
\end{equation} 
The top eigenvalue, $\lambda^1=1+\tanh{J}>1$, and the smallest eigenvalue, $\hat \lambda^1 =\lambda^N=1-\tanh J<1$, are attached to the eigenvectors
\begin{equation} 
{ \bf v}^1= \frac 1{\sqrt{2}} 
\left( \begin{array}{c}
1\\1\\0\\\vdots \\0
\end{array} \right) \ , \quad  \hat{\bf v}^1= \frac {1} {\sqrt{2}} 
\left( \begin{array}{r}
1\\-1\\0\\\vdots \\0\end{array} \right) \ .
\end{equation}
The remaining $N-2$ eigenvalues are equal to 1. Using formula~(\ref{j0}) for the lowest order coupling, $J^0$, we find that those eigenmodes do not contribute and that the interaction can take three values, depending on the choices for $p$ and $\hat p$: 
\begin{eqnarray}\label{couplC}
(J^0)_{p=1,\hat p=0} &=& \frac{\tanh{J}}{2\,(1+\tanh{J})}\simeq \frac J2 -  \frac {J^2}2 + \frac{J^3}3 + \ldots   \ , \nonumber \\
(J^0)_{p=0,\hat p=1} &=& \frac{\tanh{J}}{2\,(1-\tanh{J})}\simeq \frac J2 +  \frac {J^2}2 + \frac{J^3}3 + \ldots \ , \nonumber \\
(J^0)_{p=1,\hat p=1} &=& \frac{\tanh{J}}{1-\tanh^2{J}}\simeq J  + \frac{2\, J^3}3 + \ldots  \ .
\end{eqnarray}
Those expressions are plotted in Fig.~\ref{infj12}. The coupling $(J^0)_{1,0}$ (dashed line), corresponding to the standard Hopfield model, saturates at the value $\frac 14$ and does not diverge with $J$. Even the small $J$ behavior, $(J^0)_{1,0}\simeq \frac J2$, is erroneous. Adding the repulsive pattern leads to a visible improvement, as fluctuations of the spin configurations along the eigenvector $\bf{\hat v}^1$ (one spin up and the other down) are penalized. The inferred coupling, $(J^0)_{1,1}$ (bold line), is now correct for small $J$, $(J^0)_{1,1}\simeq J$, and diverges for large values of $J$.

\begin{figure}
\begin{center}
\vskip 1cm
\epsfig{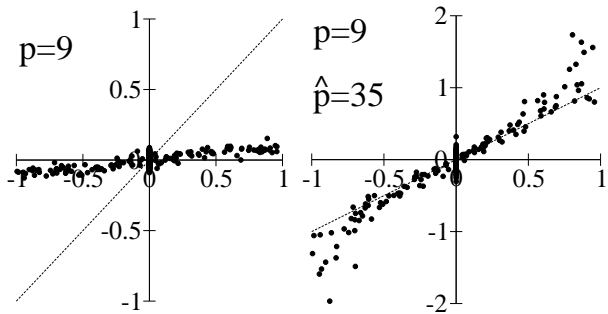}
\caption{Inferred vs. true couplings for Model D, with $B=4500$ sampled configurations. {\bf Left}: Hopfield model with $p=9$ (corresponding to the optimal number of patterns selected by the geometrical criterion); no repulsive pattern is considered $(\hat p=0)$. {\bf Right:} Generalized Hopfield model with $(p,\hat p)=(9,35)$ (optimal numbers).}
\label{fig-D2}
\end{center}
\end{figure}

We now turn to Model D. Figure \ref{fig-D2} compares the inferred and true couplings for $B=4500$ sampled configurations. The generalized Hopfield model outperforms the standard Hopfield model ($\hat p=0$), showing the importance of repulsive patterns in the inference of sparse networks with strong interactions. Large couplings, either positive or negative, are overestimated by the lowest order ML estimators for the patterns. 

\subsection{Error bars and criterion for $p,\hat p$}

An illustration of formula (\ref{errorbar}) for the error bars is shown in Fig.~\ref{fig-compar-1patt}, where we compare the components of the true pattern used to generate data in Model A with the inferred one, $(\xi ^0)_i$, and the error bar, $\sqrt{\langle (\Delta \xi_i)^2\rangle}$. For small $\alpha=\frac BN$ the inferred pattern components are uncorrelated with the true pattern and compatible with zero within the error bars. For larger values of $\alpha$, the discrepancy between the inferred and the true components are stochastic quantities of the order of the calculated error bars.

\begin{figure}
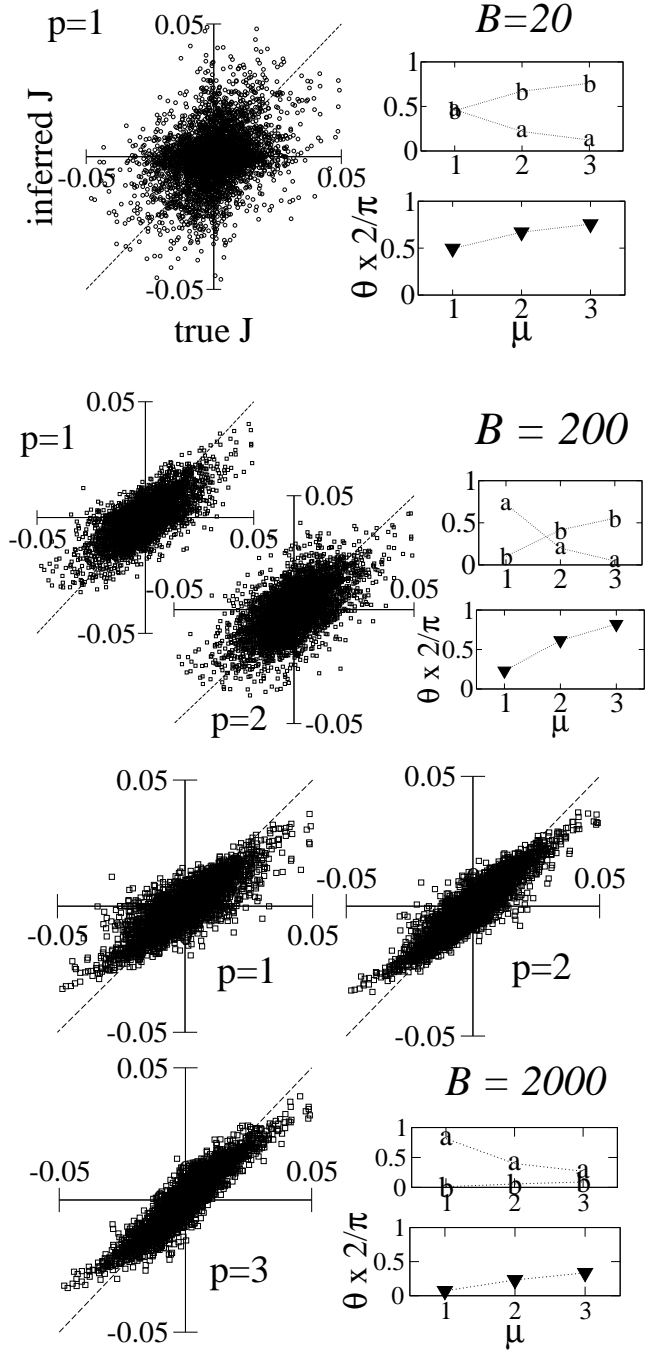

\begin{center}
\epsfig{file=./fig8a.eps,width=8.cm}\\
\vskip .5cm
\epsfig{file=./fig8b.eps,width=8.5cm}\\
\vskip .3cm
\epsfig{file=./fig8c.eps,width=8.cm}
\caption{Criterion to decide the number $p$ of patterns and performance of the ML inference procedure for three different sizes of the data set, $B$. {\bf Left:} inferred vs. true interactions with $p=1,2$ or 3 patterns; the dashed line has slope unity. {\bf  Right:} coefficients $a^\mu=(\rho^\mu)^2$ and $b^\mu=\langle ({\boldsymbol \beta}^\mu)^2\rangle$ vs. pattern index $\mu$, and  angles $\theta^\mu$, divided by $\frac{\pi}2$, see definitions (\ref{fluc78}) and (\ref{gc}). For each value of $B$ one data set was generated from Model A with $p=3$ patterns, and standard deviations $\xi^1=.95$, $\xi^2=.83$,  and $\xi^3=.77$. }
\label{fig-optimalp}
\end{center}
\end{figure}

We report in Fig.~\ref{fig-optimalp} the tests of the criterion for determining $p$ and $\hat p$  on artificially generated data from an extension of model A with $p=3$ patterns. For very poor sampling (Fig.~\ref{fig-optimalp}, top) the angle $\theta^1$ is close to $\frac{\pi}4$: even the first pattern cannot be inferred correctly. This prediction is confirmed by the very poor comparison of the true interactions and the inferred couplings calculated from the first inferred pattern. For moderately accurate sampling (Fig.~\ref{fig-optimalp}, middle) the strongest pattern can be inferred; the accuracy on the inferred couplings worsens when the second pattern is added. Excellent sampling allows for a good inference of the structure of the underlying model: the angle $\theta^\mu$ is small for $\mu=1,2,3$ (Fig.~\ref{fig-optimalp}, bottom), and larger than $\frac{\pi}4$ for $\mu\ge 4$ (not shown). Not surprisingly large couplings are systematically affected by errors. Those errors can be corrected by taking into account $O(\frac {\xi}{\sqrt N})$ corrections to the patterns if the number of data, $B$ , is large enough (Section \ref{size}).

\begin{figure}
\begin{center}
\epsfig{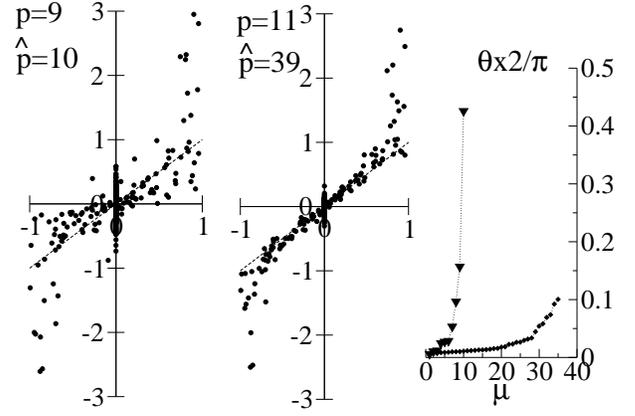}
\caption{Inferred vs. true couplings for Model D, with $B=4500$ sampled configurations.  {\bf Left:} Generalized Hopfield model with $(p,\hat p)=(9,10)$ and  $(11,39)$ (corresponding to the numbers of eigenvalues, respectively, larger and smaller than unity). {\bf Right:} angles $\theta^\mu$ and $\hat \theta ^\mu$ for, respectively, attractive (triangle) and repulsive (diamond) patterns.}
\label{fig-D1}
\end{center}
\end{figure}
        
Figure \ref{fig-D1} compares the inferred and true couplings for $B=4500$ sampled configurations of Model D. The optimal number of patterns given by the geometrical criterion is ($p=9,\hat p=35$), see Fig.~\ref{fig-D2}. Hence most of the components of $\Gamma$ are retained and the interactions inferred with the generalized Hopfield model do not differ much from the MF couplings. 

\subsection{Corrections to the patterns}\label{seccorrsyn}

\begin{figure}
\begin{center}
\epsfig{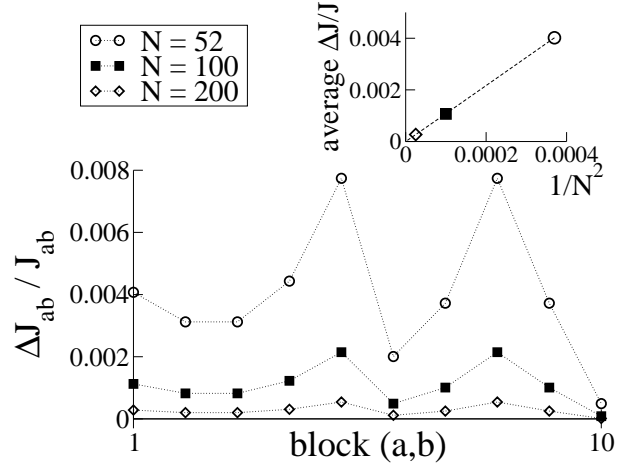}
\caption{Relative differences between the true and the inferred couplings, $\Delta J_{ab}/J_{ab}$ as a function of the system size, $N$. The inference was done using the finite--$N$ ML formulae (\ref{ordre0xi}) and (\ref{ordre1xi}) for the patterns. Data were generated from a perfect sampling of the equilibrium distribution of a Hopfield model with $p=3$ patterns and four blocks of $\frac N4$ spins, see main text; $a$ and $b$ are the block indices.  Inset: average value of $\Delta J_{ab}/J_{ab}$ as a function of $\frac 1{N^2}$. Circles, squares and diamonds correspond to, respectively, $N=52$, 100 and 200 spins.}
\label{fig-compar-3patt-corr}
\end{center}
\end{figure}

Formula (\ref{ordre1xi}) for the corrections to the patterns was tested on model B in the case of perfect sampling. Results are reported in Fig.~\ref{fig-compar-3patt-corr} and show that the errors in the inferred couplings are much smaller than in Fig.~\ref{fig-compar-3patt}. Inset of Fig.~\ref{fig-compar-3patt-corr} shows that the relative errors are of the order of $\frac 1{N^2}$ only. This scaling was expected from Section \ref{secqual}. Pushing our expansion of $\xi$ to the next order in powers of $\frac 1N$ could in principle give explicit expressions for those corrections. We have also tested our higher order formula when the fields $h_i$ are non-zero. For instance we have considered the same Hopfield model with $p=3$ patterns as above, and with block pseudo-magnetizations ${\bf t}=\frac 1{15}(2\sqrt 3,2,2,-4)$. Hence, $\bf t$ was orthogonal to the patterns, and the field components were simply given by $h_i=\tanh^{-1} t_i$, according to (\ref{changevar}) \footnote{The corresponding magnetizations were $\simeq (-.26,.13,.13,.23)$ for $N=52$ spins.}. For $N=52$ spins the relative error over the pseudo-magnetizations (averaged over the four blocks $a$) was $\frac{\Delta t_a}{t_a}\simeq .0301$ with the large-$N$ formula (\ref{ordre0xi}) and $\frac{\Delta t_a}{t_a}\simeq 0.0029$ with the finite-$N$ formulae (\ref{ordre1xi}) and (\ref{ordre1h}).

\begin{figure}
\begin{center}
\epsfig{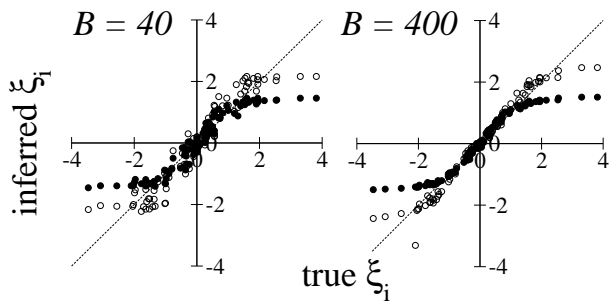}
\caption{True vs. inferred components of the patterns, $\xi_i$, for the model with $N=100$ spins described in Fig.~\ref{fig-compar-1patt-bis}. Full circles are the result of the lowest order inference formula (\ref{ordre0xi}), while empty circles show the outcome of the first order formulae (\ref{ordre1xi}).  }
\label{fig-compar-patt1-corr}
\end{center}
\end{figure}

\begin{figure}
\begin{center}
\epsfig{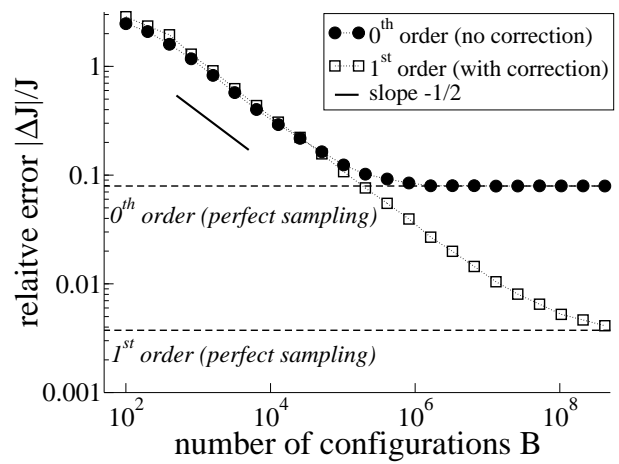}
\caption{Relative error between the inferred and true couplings for Model B (with $N=52$ spins) vs. number of sampled configurations, $B$. The two curves correspond to the inference done with the $0^{th}$ order formula (\ref{ordre0xi}) (black circles) and the $1^{st}$ order formula (\ref{ordre1xi}) (squares). Each data point is the average over 10 samples; relative error bars are about 1\%, and are much smaller than the symbol size. The asymptotic value of the errors, corresponding to perfect sampling ($B=\infty)$, are extracted from Figs.~\ref{fig-compar-3patt} and \ref{fig-compar-3patt-corr}.  }
\label{fig-compar-pca-corr}
\end{center}
\end{figure}

Corrections to the PCA were also tested when data are corrupted by sampling noise. We compare in Fig.~\ref{fig-compar-patt1-corr} the components of the pattern of Model A found with the lowest order approximation (\ref{ordre0xi}) and with our first order formulae (\ref{ordre1xi}) (case of strong pattern). A clear improvement in the quality of the inference is observed, even when the sampling noise is strong. Our second example is Model B. We show in Fig.~\ref{fig-compar-pca-corr} the relative errors 
\begin{equation}\label{epsJ}
\epsilon _J = \frac 2{N(N-1)} \sum _{i<j} \left|\frac{\Delta J_{ij}}{J_{ij}}\right|
\end{equation}
between the true and the inferred couplings, with formulas (\ref{ordre0xi}) and (\ref{ordre1xi}), as a function of the number of sampled configurations, $B$, and for $N=52$ spins. As $B$ increases the relative error with the lowest order patterns (PCA) first decreases as $B^{-1/2}$, then saturates to the value $\simeq .0794$, as expected from Fig.~\ref{fig-compar-3patt}. The relative error with the correction to the patterns also decreases as $B^{-1/2}$, and is expected to saturate to the lower value $\simeq .00374$ (Fig.~\ref{fig-compar-3patt-corr}). We remark that the gain in accuracy over the inferred couplings resulting from the corrections (\ref{ordre1xi}) to the patterns is obtained only when $B$ is very large. $B\sim N^3$ configurations at least should be sampled to obtain an improvement over the lowest order formula (\ref{ordre0xi}). This scaling holds when the couplings are weak, and decrease as $\frac 1N$. If the interaction network is diluted and carries couplings $J=O(1)$, we expect that $B\sim N/J^2$ configurations have to be sampled to make the first-corrections to the patterns effective.

We have applied our formula (\ref{ordre1xi}) to calculate the first correction to the couplings (\ref{couplC}) for Models C and D. As for Model C, we find that the correction to the coupling $(J^0)_{1,1}$ vanishes; this result is due to the fact that $(J^0)_{1,1}$ is already correct to the second order in $J$, and that higher order corrections would be needed. The corrections to the coupling $(J^0)_{1,0}$ are equal to
\begin{eqnarray}
(J^1)_{1,0}&=& \frac{\tanh{J}}{2\sqrt{2}}+\frac{\tanh{J}(1+\tanh{J})}{16} \nonumber \\
&=& \left(\frac 1{16}+\frac 1{2\sqrt 2}\right) J + \frac{J^2}{16} + \ldots ... \ .
\end{eqnarray}
The resulting coupling, $(J^0+J^1)_{10}$, is plotted as a function of $J$ in Fig.~\ref{infj12}, and qualitatively improves over the lowest order result (\ref{couplC}). In particular, for small $J$, the inferred coupling is now $(J^0+J^1)_{10}\simeq .916 \,J -.438 \, J^2$, which is definitely closer to $J$ than (\ref{couplC}). In the case of Model D, the first-order corrections improve only slightly the estimates for the large couplings. 

\section{Application to biological data}
\label{secbio}

In this Section we show how the inference approach can be applied to real biological data, and compared to other Boltzmann Machine learning procedures.

\subsection{Cortical activity of the rat}

\begin{figure}
\begin{center}
\epsfig{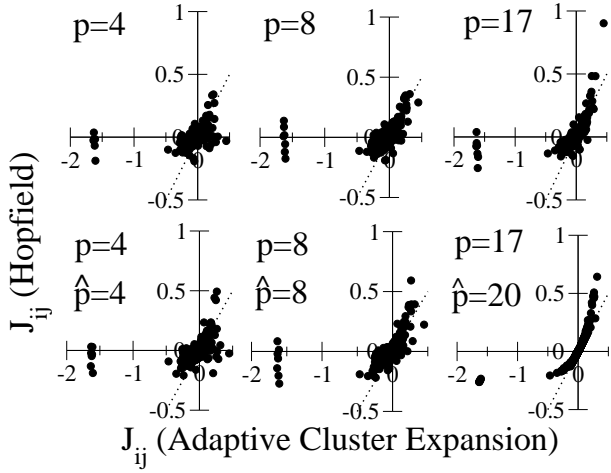}
\caption{Couplings calculated with the generalized Hopfield model vs. couplings calculated with the adaptive cluster expansion of \cite{noi} for 37 cells recorded in the prefrontal cortex of a behaving rat.  {\bf Top}: Hopfield model with $p=4$, $8$ (corresponding to the optimal number of patterns selected by the geometrical criterion) and $17$; no repulsive pattern is considered $(\hat p=0)$. {\bf Bottom.} Generalized Hopfield model with $(p,\hat p)=(4,4)$, $(8,8)$ (optimal numbers) and  $(17,20)$ (corresponding to the numbers of eigenvalues, respectively, larger and smaller than unity).}
\label{adrien}
\end{center}
\end{figure}

We have first analyzed data coming from the recording of 37 neurons in the prefrontal cortex of rats. The experiment, done by A. Peyrache, F. Battaglia and their collaborators, consists in recording the neural activity during a task and during the Slow Wave sleep preceding and following the learning of the task \cite{pey09}. PCA allowed Peyrache et al. to identify patterns in the activity, which are generated when the rat learns a task and are replayed during the sleep \cite{pey09}. 

We have analyzed with the generalized Hopfield model the data corresponding to a 20 minute-long recording of the activity of a rat during the task (data shown in Fig. 1 of ~\cite{pey09}). The raster plot was binned with a 10 msec window to obtain binary configurations of the neurons (active or silent in the time-bin). We have then calculated the average frequencies, $m_i$, and the pairwise correlations, $c_{ij}$. We calculate the couplings with $p$ attractive and $\hat{p}$ repulsive patterns according to (\ref{ordre0xi}) and (\ref{j0}). The numbers $p$ and $\hat p$ are calculated according to the geometrical criteria (\ref{gc}) and (\ref{gcr}). Hereafter, we compare the couplings obtained this way to the ones found with the adaptive cluster expansion (ACE) of \cite{noi}, which is not based on the expansion of the loglikelihood used in the present work. 

In Fig.~\ref{adrien} (top) we compare the Hopfield ($\hat p=0$) couplings with $p=4,8,17$ selected patterns to the ACE couplings. The agreement is quite good for $p^*=8$. In \cite{pey09} $p=6$ patterns were kept in the PCA; this value is close to the optimal value, $p= 8$, we find using the geometrical criterion. Addition of repulsive patterns (bottom of Fig.~\ref{adrien}) slightly improves the similarity with the ACE couplings. We find, indeed, that the couplings $J_{ij}$ are rather weak, and that repulsive patterns do not play an important role. Calculating the couplings with all eigenmodes ($p=17,\hat p=20$) is equivalent ot the mean-field (MF) approximation. A clear discrepancy between the Hopfield and the ACE couplings is found for the largest (in absolute value) interactions. We have checked that this discrepancy is not reduced when the first order corrections to the patterns are included, presumably because the number of data is not sufficient. Couplings are not significatively changed in the presence of the regularization (\ref{regu1}) for sensible values of $\gamma$.
 
\subsection{Protein-domain families}

\begin{figure}
\begin{center}
\epsfig{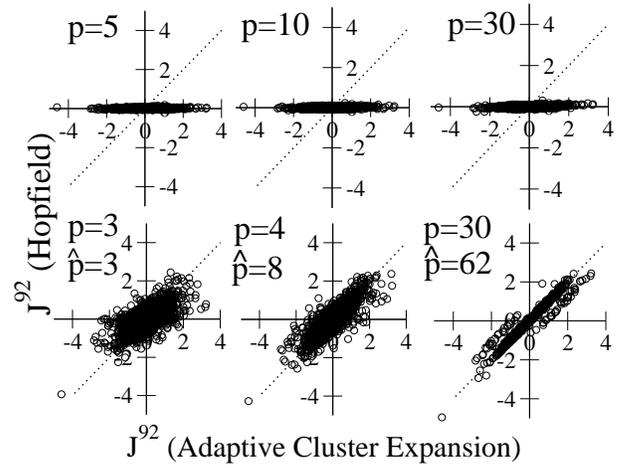}
\caption{Couplings calculated with the Generalized Hopfield model versus coupling calculated with the adaptive cluster expansion for 92 amino-acids in the PDZ domain.The values of $p,\hat p$ are given in the Figure. Note that $\hat p=0$ for the top panels. The middle panels correspond to the optimal values for the number of patterns.}
\label{jpdz}
\end{center}
\end{figure}

We have next analyzed the alignement of a family of 240 sequences of PDZ, a commonly encountered domain binding the C-terminus of proteins, with $92$ amino-acids \cite{ran}. R. Ranganathan and collaborators have elaborated an approach, called Statistical Coupling Analysis(SCA), to extract interactions between residues by using evolutionary data for the protein, {\em i.e.} by sampling the single-site and pairwise frequencies from multi-sequence alignments of the family \cite{loc99}.  Briefly speaking, SCA consists in doing a PCA analysis of a weighted correlation matrix, $D_i\Gamma_{ij} D_j$, where the weight $D_i$ on site $i$ is small for poorly conserved residues \cite{ran}.  

We have taken the binary data representation of the 240 PDZ sequences in the alignement given in~\cite{riv09} (Supplementary Material). This consensus approximation amounts to replace the amino-acid on each site (20 possible types) with a binary variable $\sigma_i^b$, equal to $+1$ if the amino-acid $i$ in the $b^{th}$ sequence is the most common amino-acid at that position in the alignment, to $-1$ otherwise. The consensus representation does not allow to keep track of all the information contained in the alignment but is indicative of the conservation pattern in the family.

The inferred couplings, denoted by $J^{92}$, are shown in Fig.~\ref{jpdz}. As in the case of Model D in Section \ref{secsyn} we find that proteomic data are better accounted with the generalized Hopfield model than with the standard Hopfield model: repulsive patterns seems necessary to recover the couplings found with the ACE method. The couplings found with attractive patterns only are not correlated with the ACE couplings (top of Fig.~\ref{jpdz}), while the agreement is quite good when taking into account attractive and repulsive patterns; the optimal numbers of  patterns are $p=4$ and $\hat p=10$.

\begin{figure}
\begin{center}
\epsfig{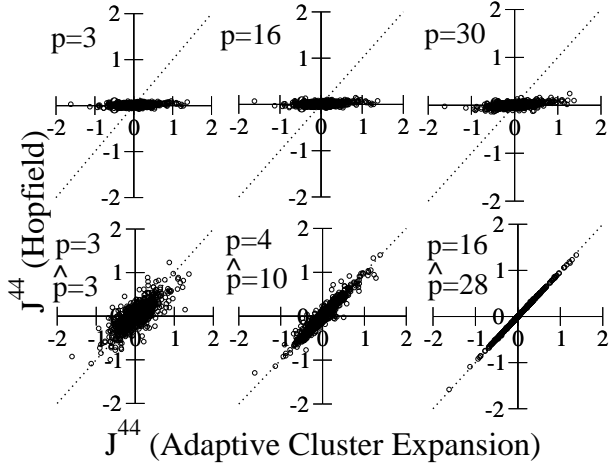}
\caption{Same as Fig.~\ref{jpdz} when retaining the 44 residues with the largest weights $D_i$ only \cite{ran}. The values of $p,\hat p$ are given in the Figure. Note that $\hat p=0$ for the top panels. The middle panels correspond to the optimal values for the number of patterns.} 
\label{jvsjpdz44}
\end{center}
\end{figure}

We have also calculated the couplings when discarding all but the most weighted sites. More precisely, we have recalculated the distribution of the weights $D_i$ as in \cite{ran,riv09}, and found a bimodal distribution, which suggests a natural cut-off between large and small weights. We have redone the previous inference when keeping only the 44 residues (out of 92) with the largest weights, corresponding to the red sites in Fig.~C of \cite{ran}. The resulting interactions, denoted by $J^{44}$, are shown in Fig.~\ref{jvsjpdz44}. Again we compare the couplings found with the Hopfield model and with the ACE. The agreement is not good with attractive patterns only (as done in usual PCA), and is very good when repulsive patterns are included. 

\begin{figure}
\begin{center}
\epsfig{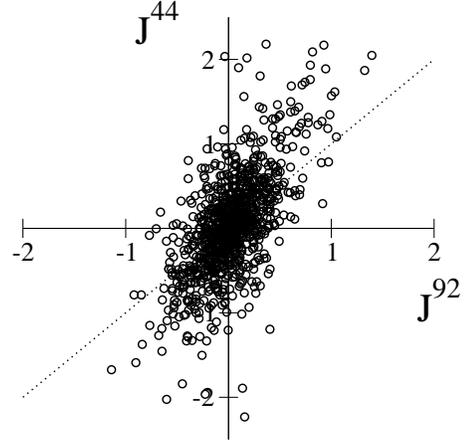}
\caption{Comparison between the couplings $J_{ij}$ calculated with all $92$ residues and with the $44$ most weighted residues only, for each one of the $44\times 43/2$ pairs $(i,j)$ of residues.}\label{j92vsj44}
\end{center}
\end{figure}

An interesting question is whether the couplings obtained between the 44 most conserved residues are strongly affected by the presence or the absence of the remaining 48 residues in the inference. The interactions in the $44$-site model are effective and {\em a priori} differ from their values in the $92$-site model, in that they account for chains of interactions going through the remaining 48 sites. Nevertheless, we find that the couplings calculated with all 92 residues and the couplings obtained from the subset of 44 sites with large weights are similar, see Fig.\ref{j92vsj44}. This result suggests that the 48 residues removed from our second analysis are not strongly interacting with the 44 retained sites. 

\section{Expansion of the cross entropy and maximum likelihood inference}
\label{secinf}

This Section is intended to provide the derivations of the results announced in Section \ref{mainres}. Maximizing the posterior probability (\ref{post}) with respect to the patterns and the fields is equivalent to minimizing the cross entropy of the Hopfield model given the data,
\begin{equation}
\Phi [{\bf h},\{ \boldsymbol \xi^\mu\},\{\boldsymbol \sigma ^b\}]=\log{Z} 
[{\bf h},\{ \boldsymbol \xi^\mu\}] +U [{\bf h},\{ \boldsymbol \xi^\mu\},\{\boldsymbol \sigma ^b\}]\ ,
\label{Fi}
\end{equation}
where $Z$ is the partition function appearing in (\ref{likelihood}),
\begin{equation}\label{pf}
Z[{\bf h},\{ \boldsymbol \xi^\mu\}] =\sum_{\boldsymbol\sigma} \exp \big( - E[\boldsymbol\sigma, {\bf h},  \{\boldsymbol \xi^\mu\}]\big)\ ,
\end{equation} 
and $U$ is the average value of the energy $E$ (\ref{energy}) over the sampled configurations:
\begin{equation}
U[{\bf h},\{ \boldsymbol \xi^\mu\},\{\boldsymbol \sigma ^b\}]= -\sum_{i=1}^N  h_i m_{i}-\frac 1{2} \sum_{i,j} J_{ij} \, c_{ij}\ , 
\label{secexpa}
\end{equation}
where the couplings $J_{ij}$ are calculated from the patterns according to (\ref{defcoupl}).
The calculation of the partition function, which is defined as a sum over $2^N$ configurations, cannot generally be done in a reasonable time for large sizes $N$. In the next section we show how the use of statistical mechanics techniques allows one to obtain a systematic expansion of $Z$, and, thus, of the cross entropy
\begin{equation}\label{expphi}
\Phi=\Phi^0+\Phi^1 +\ldots \ ,
\end{equation}
in powers on $\frac {\xi _{i}}{\sqrt N}$ and $\frac {\hat \xi_{i}}{\sqrt N}$.

\subsection{Expansion of the free energy of the  Hopfield model in powers of $\frac {\xi_{i}}{\sqrt N},\frac {\hat \xi_{i}}{\sqrt N}$}\label{exp67}

To lighten notations calculations are presented for the case of attractive patterns only. We explain at the end of the Section how formulae are modified in the presence of repulsive patterns.

For technical reasons to be made clear below it results convenient to make the change of variables ${\bf h}\to{\bf t}$ described by
\begin{eqnarray}\label{changevar}
h_i &=& \tanh^{-1} t_i - \frac 1N \sum _\mu \sum_j \xi_i^\mu  \xi_j^\mu \, t_j \ ,
\end{eqnarray}
where the $t_i$, hereafter called pseudo-magnetizations, are real-valued numbers comprised between $-1$ and $1$. 
Hereafter, we will infer the most likely values for $\bf{t}$, and will recover the fields $\bf h$ through (\ref{changevar}). The change ${\bf h}\to{\bf t}$  amounts to consider the energy function
\begin{equation}\label{energy2}
E=-\sum_{i=1}^N \sigma _{i} \tanh^{-1} t_i- \frac 1{2N} \sum_{\mu=1}^p \left( \sum_{i=1}^N \xi_{i}^\mu\big(\sigma_{i}-t_i\big)\right)^2 \ ,
\end{equation}
instead of the original expression for $E$ (\ref{energy}) (with $\hat p=0$). Obviously, when the identities (\ref{changevar}) are fulfilled, both energies are equal (up to a $\boldsymbol\sigma$-independent additive term) and define the same likelihood function (\ref{likelihood}). 

We unravel the squared terms in the partition function (\ref{pf}) through a set of $p$ auxiliary Gaussian variables ${\bf {x}}=( {x}^1,\ldots,{x}^p)$, and carry out the summation over the spin configurations. We obtain
\begin{eqnarray}\label{zz}
Z &=& \int \prod_{\mu}\frac{dx^\mu}{\sqrt {2\pi}}\; \exp\bigg[ -
\frac 12 \sum _{\mu} (x^\mu)^2 -\sum_{i,\mu} \frac {x^\mu\, \xi_i^\mu\,t_{i} }{\sqrt N} \nonumber \\
&+&\!\! \sum_{i} \log 2 \cosh \left( \tanh^{-1} t_{i} + \sum_{\mu} \frac {x^\mu\, \xi_i^\mu }{\sqrt N}\right) \bigg]\; .
\end{eqnarray}
If $N$ is large enough the dominant contribution to the integral will come from ${\bf x}^*$, the value of ${\bf x}$ maximizing the argument of the exponential above. We obtain the following saddle point equation for ${\bf x}$,
\begin{equation}
({x^{\mu}})^* = \frac{1}{\sqrt {N}} \sum_i \xi_i^\mu (T_i-t_i) \ ,
\label{xmu}
\end{equation}
where
\begin{equation}\label{deftig}
T_i\equiv \tanh\left( \tanh^{-1} t_{i} + \sum_{\mu} \frac {({ x^\mu})^*\, \xi_i^\mu }{\sqrt N}\right)
\end{equation}
We then write $x^\mu= ( x^\mu)^* + y^\mu $ and expand the hyperbolic cosine function in powers of $y^{\mu}$. The change of variable (\ref{changevar}) is such that the linear term  in $y^{\mu} $ in the expansion of the hyperbolic cosine function cancels out with the linear term in the exponential, $\displaystyle{-\sum_{i,\mu} \frac {y^\mu\, \xi_i^\mu\,t_{i} }{\sqrt N}}$, independently of the value of $(x^{\mu})^*$. Expanding the hyperbolic cosine up to the second order in  $y^\mu$ we find our lowest order approximation to the partition function,
\begin{eqnarray}
Z^0 &=& e^{F^*}\,\int \prod_{\mu} \frac{d {y}^\mu}
{\sqrt {2\pi}}\; \exp\bigg[- \frac 12 \sum_{\mu}  ({y}^{\mu})^2 
\\&+&\frac{1}{2\,N} \sum_i \sum_{\mu,\nu} \xi_i^\mu  \xi_i^\nu
 {y}^{\mu}  {y}^{\nu}  \left(1- T_i^2\right) \bigg]  =  \frac {e^{F^*}}{\sqrt{\hbox{\rm det A}}} \nonumber 
\label{Z0}
\end{eqnarray}
where $F^*$ is the the argument of the exponential in (\ref{zz}) calculated in  ${x^\mu}^*$,
\begin{equation}
F^*=N\log 2+ \frac 12 \sum _i \log (  1-T_i^2) -\sum_{\mu,ij} \xi_i^\mu \xi_j^\mu (T_i\,T_j-t_i\,t_j)\ ,
\end{equation}
and $A$ is the $p\times p$ matrix with entries,
\begin{equation}\label{matrixa}
A^{\mu\nu} = \delta ^{\mu\nu}- \frac 1N \sum _i \xi_i^\mu\xi_i^\nu \big( 1- T_i^2\big)\ .
\end{equation}
We then compute the average energy $U$ (\ref{secexpa}), 
\begin{eqnarray}
U&=& -\sum_{i}   m_{i} \tanh^{-1}\,t_i \label{u}\\ &-&
\frac 1{2N} \sum_{\mu,ij} \xi_{i}^\mu \xi_{j}^\mu \,(c_{ij}-m_i\,t_j-t_im_j+t_i\,t_j) \ . \nonumber
\end{eqnarray}
Our lowest order approximation for the cross entropy is, according to (\ref{Fi}), (\ref{Z0}) and (\ref{u}):
\begin{eqnarray}\label{logppost0}
\Phi^0&=& -\sum_{i=1}^N   m_{i} \tanh^{-1}T_i+ N\log 2+ 
\frac 12 \sum _i \log (  1-T_i^2) \nonumber \\  
&-& \frac 1{2N}\sum_{\mu,ij} \xi_i^\mu \big(c_{ij}-m_i\,m_j\big)\xi_j^\mu -
\frac 12 \log\det A \nonumber \\&+& \frac 1{2N} \sum _\mu \left[ \sum_i \xi_i^\mu\big(T_i-m_i\big)\right]^2\ .
\end{eqnarray} 
The first order contribution to the cross entropy, $\Phi^1$ in (\ref{expphi}), is obtained by retaining the fourth order in $y^\mu$ in the expansion of the hyperbolic cosine function in (\ref{zz}),
\begin{equation}\label{logppost1}
\Phi^1=\frac 1{4N^2} \sum_i (1-4T_i^2+3T_i^4) \bigg( \sum _{\mu,\nu} \xi_i^\mu (A^{-1})^{\mu\nu} \xi_i^\nu \bigg)^2 \; .
\end{equation}
We expect the differences $\Phi-\Phi^0$ and $\Phi-(\Phi^0+\Phi^1)$ between, respectively, the true and the lowest order cross entropies and the true and the first order cross entropies to be of the order of, respectively, $R^2$ and $R^3$, where 
\begin{equation}\label{expr}
R =\frac pN\, \xi^2 \, (1-m^2)\, \Lambda   \ .
\end{equation}
Here, $\xi^2$ is the order of magnitude of the pattern components, which can range from 1 if the patterns are extended over the whole system to $\sim \sqrt N$ for highly sparse patterns, $m$ is the typical value of the local magnetization, and $\Lambda$ is the order of magnitude of the eigenvalues of $A^{-1}$, which can range from 1 to $N$. The value of $R$ fixes the instrinsic error $\epsilon$ on the inferred patterns discussed in Section \ref{secqual}, $\epsilon \sim R$ for the lowest order approximation and $\epsilon \sim R^2$ with the first order corrections.
 
The above calculation can be straightforwardly extended to the case of the generalized Hopfield model by considering the $\hat p$ repulsive patterns as patterns with purely imaginary components, $\boldsymbol\xi^{\mu} = i \,\hat {\boldsymbol\xi}^{\mu}$, with $i^2=-1$. For instance the general lowest order expression for the cross entropy is     
\begin{eqnarray}\label{logppost0b}
\Phi^0&=& -\sum_{i=1}^N   m_{i} \tanh^{-1}T_i+ N\log 2+ 
\frac 12 \sum _i \log (  1-T_i^2) \nonumber \\  
&-& \frac 1{2N}\sum_{ij}\big(c_{ij}-m_i\,m_j\big)\left(\sum _{\mu=1}^p \xi_i^\mu \xi_j^\mu -\sum _{\mu=1}^{\hat p} \hat\xi_i^\mu \hat\xi_j^\mu\right)  \nonumber \\
&+& \frac 1{2N} \sum _{\mu=1}^p \left[ \sum_i \xi_i^\mu\big(T_i-m_i\big)\right]^2  \nonumber \\ &-&\frac 1{2N} \sum _{\mu=1}^{\hat p} \left[ \sum_i \hat{\xi}_i^\mu\big(T_i-m_i\big)\right]^2  \nonumber \\
&-&\frac 12 \log \det \left( \begin{array} {c c}
A & i \hat A\\
- i {\hat A}^T & \hat{\hat A} 
\end{array}\right)\ ,
\end{eqnarray} 
where
\begin{eqnarray}\label{matrixab}
 T_i \!\!&=& \!\!\tanh\!\left(\! \tanh^{-1} t_{i} + \sum_{\mu=1}^p \frac 
{ (x^\mu)^* \xi_i^\mu}{\sqrt{N}} - \sum_{\mu=1}^{\hat p} \frac {( {\hat x}^{\mu})^*\hat\xi_i^\mu }{\sqrt N}\right) \, ,\nonumber \\
(\hat x^{\mu})^*\!\!\!&=& \frac{1}{\sqrt {N}} \sum_i \hat \xi_i^\mu (T_i-t_i) \ ,\nonumber \\ 
\hat A^{\mu\nu} &=&  \frac 1N \sum _i \xi_i^\mu\hat{\xi}_i^\nu \big( 1- T_i^2\big)\ ,\nonumber \\ 
\hat{\hat A}^{\mu\nu} &=& \delta ^{\mu\nu}+ \frac 1N \sum _i 
\hat{\xi}_i^\mu\hat{\xi}_i^\nu \big( 1- T_i^2\big)\ .
\end{eqnarray}
The first order correction (\ref{logppost1}) can be easily written for the case of repulsive patterns, too.

\subsection{Are the physical properties of the system relevant for the inference?}
\label{seccoex}
The Hopfield model was first introduced as a model for which a set of $p$ desired ground states $\boldsymbol\xi^\mu$ (or fixed points of the zero temperature Glauber dynamics) could be programmed through an adequate choice of the interactions. Each fixed point has a basin of attraction in the configuration space, corresponding to a phase of the system. The order parameters are the overlaps
\begin{equation}\label{defqmu}
q^\mu=\sum _{\boldsymbol\sigma} P_H[\boldsymbol\sigma | {\bf h}, \boldsymbol \xi]  \ \left( \frac 1N \sum _i \xi_i^\mu \,\sigma _i \right) \ ,
\end{equation}
which quantify how much the configurations are on average aligned along each pattern. The amplitudes and directions of the pattern and the field vectors determine if spin configurations tend to be aligned along the field, or along one or more patterns. In the infinite size limit ($N\to\infty$) the overlaps are the roots of $p$ coupled and self-consistent equations,
\begin{equation}\label{scm}
q^\mu =\lim _{N\to\infty}\frac 1N \sum _i \xi_i^\mu\,  \tanh \big(  h_i + \sum_\rho q^\rho \xi_i^\rho \big) \ .
\end{equation}
Using (\ref{changevar}) and  the saddle point equation (\ref{xmu}) it is easy to check that the overlaps 
\begin{equation}\label{sol}
q^\mu=\frac 1N \sum _i \xi_i^\mu\; T_i 
\end{equation}
are solutions to the set of equations (\ref{scm}). Solutions are in one-to-one correspondance with the saddle points $(x^\mu)^*$.

The saddle-point solution ${\bf x}^*=0$ corresponds to $T_i=t_i$. The average interaction term in the energy function (\ref{energy2}) vanishes, meaning that configurations tend to be mainly determined by the fields. Such a behaviour corresponds to the paramagnetic phase. The solution ${\bf x}=0$ is locally stable if the eigenvalues of the matrix $A$ are all positive and, thus, if the patterns are weak enough. Solutions with ${\bf x}^*\neq 0$ correspond to stronger patterns and interaction terms in (\ref{energy2}) having non zero values on average: they correspond to magnetized phases.

The cross entropy $\Phi$ depends on the solution ${\bf x}^*$ through the variables $T_i$ only. Once the $T_i$'s and the patterns $\boldsymbol\xi^\mu$'s are inferred, it is easy to calculate the value of the fields $h_i$ based on equations (\ref{changevar}), (\ref{xmu}) and (\ref{deftig}). One finds that $h_i$ is given by (\ref{changevar}) where $t_i$ is substituted with $T_i$. Hence, the inferred parameters do not explicitely depend on the value of $x^*$. The procedure followed to infer the patterns and the fields is not affected by the physical phase (paramagnetic or magnetized) of the system, though the values of the data $m_i$ and $c_{ij}$ obviously depend on those physical properties. 

It may accidentally happen that equations (\ref{xmu}) have different solutions with equal or almost equal contributions to the partition function $Z$. The most natural illustration is the case of zero field ($t_i=0$) and one strong pattern, where two ferromagnetic states with opposite overlaps, $(x^1)^*$ and $-(x^1)^*$, coexist. In this latter case both states give equal contributions to the partition function. 

\subsection{Maximum Likelihood inference: lowest order}\label{secml}

We first infer the patterns and the pseudo-magnetizations from $\Phi^0$. Minimization of $\Phi^0$ (\ref{logppost0}) over ${\bf T}$ immediately shows that, up to $O(R)$ corrections, pseudo- and true magnetizations coincide: 
\begin{equation}\label{ordre0T}
(T_i)^* = m_i\ .
\end{equation}
Without loss of generality we may write the patterns to infer as
\begin{eqnarray}\label{expressxi}
\left(\xi^0\right)^\mu _i &=& \frac{\sqrt {N\,a^\mu} \; v_i^\mu + \sqrt N \beta^\mu _i}{\sqrt{1-m_i^2}}\ ,
\nonumber \\
\left(\hat \xi^0\right)^\mu _i &=& \frac{\sqrt {N\,\hat a^\mu} \; \hat v_i^\mu +\sqrt N \hat \beta^\mu _i}{\sqrt{1-m_i^2}}\ ,
\end{eqnarray}
where $\sqrt{a^\mu},\sqrt{\hat a^\mu}$ are real-valued coefficients, and ${\bf v}^\mu$ and $\hat {\bf v}^\mu$ are eigenvectors of $\Gamma$. According to identity (\ref{ordre0T}) the conditions (\ref{gauge}) are fulfilled  in the large $N$ limit if the ($p+\hat p$) vectors ${\boldsymbol \beta}^\mu$ and $\hat{\boldsymbol \beta}^\nu$ are orthogonal to each other, and to all the patterns $({\boldsymbol \xi^0})^\nu $ and $(\hat{\boldsymbol \xi^0})^\nu$. The matrices $A$ (\ref{matrixa}) and $\hat{\hat A}$ (\ref{matrixab}) are then diagonal, while $\hat A$ vanishes. We rewrite the cross entropy (\ref{logppost0b}) as
\begin{eqnarray}\label{logppost2}
\Phi^0&=&- \sum _{i}\sum_{\sigma=\pm 1}\left(  \frac{1+\sigma\,m_i}2\right) \log \left( \frac{1+\sigma\, m_i}2\right)  \nonumber \\
&-&  \frac 12\sum _\mu \lambda^\mu \, a^\mu - \frac 1{2}\sum_{ij,\mu} \beta_i^\mu\, \Gamma^{(r)}_{ij}\,\beta_j^\mu \ ,\nonumber \\
&+&  \frac 12\sum _\mu \hat{\lambda}^\mu \, \hat {a}^\mu + \frac 1{2}\sum_{ij,\mu} \hat{\beta}_i^\mu\, \Gamma^{(r)}_{ij}\,\hat{\beta}_j^\mu \ ,\nonumber \\
&-& \frac 1{2} \sum _\mu \log \left[ 1 -a^\mu- \sum _i (\beta_i^\mu)^2\right]
  \nonumber \\
&-& \frac 1{2} \sum _\mu \log \left[ 1 +\hat{a}^\mu+  \sum _i (\hat{\beta}_i^\mu)^2\right]
\end{eqnarray} 
where $\Gamma^{(r)}$ is the restriction of $\Gamma$ to the $(N-p-\hat p)$--dimensional subspace orthogonal to the $p$ largest and $\hat p$ smallest eigenvectors: 
\begin{equation}
\Gamma ^{(r)}_{ij}=\sum _{k=p+1}^{N-\hat p} \lambda^k v_i^k v_j^k \ .
\end{equation}
Minimizing $\Phi^0$ over the coefficients $a^\mu$ and the vectors ${\boldsymbol \beta}^\mu$ gives the coupled set of equations
\begin{eqnarray}
\lambda^\mu  &=& \frac {1}{1 -a^\mu-b^\mu}\ , \label{map1} \\
\sum _{j} \Gamma^{(r)}_{ij}\, \beta _j^\mu &=&   \frac {\beta_i^\mu}{1 -a^\mu-b^\mu} \label{map2}\ ,
\end{eqnarray}
where $b^\mu=({\boldsymbol\beta}^\mu)^2$ is the squared norm of ${\boldsymbol \beta}^\mu$. If the vector ${\boldsymbol \beta}^\mu$ were non zero, it would be an eigenvector of $\Gamma$ with eigenvalue $\lambda^\mu$ according to (\ref{map2}). This cannot be true as the largest eigenvalue of $\Gamma ^{(r)}$ is smaller than $\lambda^p$. Hence, ${\boldsymbol\beta}^\mu=b^\mu=0$. From (\ref{map1}) we obtain 
\begin{equation} \label{amumap1}
a^\mu= 1-\frac 1{\lambda^\mu} \ . 
\end{equation}
We conclude that the maximum likelihood values for the $p$ attractive patterns are given by (\ref{ordre0xi}). The minimization of $\Phi^0$ over the coefficients $\hat a^\mu$ and the vectors $\hat {\boldsymbol \beta}^\mu$ can be done along the same lines. We find 
\begin{equation} \label{amumap1b}
\hat a^\mu= \frac 1{\hat\lambda^\mu}-1 \ . 
\end{equation}
and $\hat \beta^\mu=0$. The maximum likelihood estimators for the $\hat p$ repulsive patterns are given by (\ref{ordre0xi}) again. Once the patterns are computed the values of the local fields $h_i$ are obtained from (\ref{ordre0hbis}). 

Notice that $v_i^\mu,\hat v_i^\mu$ are typically of the order of $N^{-\frac 12}$, which entails that the components of the patterns are of the order of unity. Though keeping each $\xi_{i},\hat \xi_i$ of the order of unity is a natural scaling in the infinite size limit $N\to\infty$, other scalings are possible. Consider a pair of strongly coupled spins, {\em i.e.} such that the correlation $\Gamma_{ij}$ is sizeably larger than $\frac 1N$. According to expression (\ref{defcoupl}) for the coupling $J_{ij}$ induced by the patterns between spins $i$ and $j$, we expect the pattern components to be of the order of $\sqrt N$. There is thus no compelling reason to assume that $\frac{\xi_{i}}{\sqrt N},\frac{\hat \xi_{i}}{\sqrt N}$ is vanishingly small for all components $i$. 

To end with we compute the decrease in cross entropy when adding a pattern attached to the eigenvalue $\lambda \ (=\lambda ^\mu$ or $\hat\lambda^\mu)$. Inserting expressions (\ref{amumap1},\ref{amumap1b}) for $a^\mu,\hat a^\mu$ in (\ref{logppost2}) we obtain
\begin{equation}\label{gainphi}
\Delta \Phi  = -\frac 12 \big( \lambda -1 -\log \lambda \big) \ ,
\end{equation}
a quantity which is strictly negative for $\lambda \ne 1$. Not surprisingly, adding more parameters to the model allows for a better fit of the data. We will see in Section \ref{secoptim} how the values of $p$ and $\hat p$ can be determined.

\subsection{Error bars on the patterns and fields} \label{secerrcal}

When the sample size $B$ is large the posterior distribution $P$ tends to a Gaussian law centered in the most likely values for the patterns, $\{\boldsymbol\xi^\mu\},\{\hat{\boldsymbol\xi}^\mu\}$, and the pseudo-magnetizations, ${\bf T}$. For the sake of simplicty we consider below the case of attractive patterns only; repulsive patterns can formally be seen as purely imaginary attractive patterns, see Section \ref{exp67}. Let ${\bf H}$ denote the Hessian matrix of $\Phi^0$. We find, to the leading orders,
\begin{eqnarray}
({\bf H}^{tt})_{ij} & \equiv & \frac{\partial^2  \Phi^0}{\partial T^1_i\partial T^1_j} = \frac{\delta _{ij}}{1-m_i^2} - (J^0)_{ij} \ ,\nonumber \\
({\bf H}^{\xi\xi})_{ij}^{\mu\nu} & \equiv & \frac{\partial^2\Phi^0 }{\partial  (\xi^0)^\mu_i\partial  (\xi^0)^\nu_j} = \frac{\delta^{\mu\nu}}N \bigg[m_im_j- c_{ij} +(1-m_i^2)\nonumber \\
&\times&\lambda^\mu \left(\delta_{ij} +(1-m_j^2) \sum_\rho \frac{\lambda^\rho}N (\xi^0)_i^\rho (\xi^0)_j^\rho \right)\bigg] \nonumber \\
&+&\frac{\lambda^\mu\lambda^\nu}{N^2}(1-m_i^2)(1-m_j^2) (\xi^0)_i^\nu (\xi^0)_j^\mu \ ,\\
({\bf H}^{t\xi})_{ij}^{\nu} & \equiv & \frac{\partial^2\Phi^0 }{\partial T_i\partial  (\xi^0)^\nu_j} \simeq 0 \ .
\end{eqnarray}
Here, $\delta$ denotes the Kronecker function and the expression of the lowest order coupling matrix, $J^0$, is given in (\ref{j0}). The sum over $\rho$ runs over all pattern indices. The cross second derivative, ${\bf H}^{t\xi}$, of the order of $\frac{|\xi|}N$, is much smaller than the expected order, $\frac{|\xi|}{\sqrt N}$, and can be neglected.

The covariance matrix of the Gaussian posterior probability $P$ is the inverse matrix of $B\, {\bf H}$. The inverse is properly defined in the subspace of dimension $N(p+\hat p+1)-\frac 12 (p+\hat p)(p+\hat p-1)$, orthogonal to the modes generating the invariance over the patterns, see Section \ref{secgenhop}. We write $\tilde {\bf H}=D {\bf H} D$, where $D$ is a diagonal matrix with elements: $D_i=\sqrt{1-m_i^2}$ in the ${\bf T}$-sector, and $D_i^\mu= \sqrt{\frac{N}{1-m_i^2}}$ in the ${\boldsymbol\xi}^\mu$-sector. Matrix $\tilde{\bf H}$ has a particularly simple expression in the eigenbasis of the correlation matrix $\Gamma$, and can be diagonalized exactly after some simple algebra. We obtain the following expression for the covariance matrix of the fluctuations:
\begin{equation}
\langle \Delta T_i \,\Delta T_j\rangle= \frac{\sqrt{(1-m_i^2)(1-m_j^2)}}{B}\; \big[{\bf M}^{tt}\big] _{ij}\ , 
\label{errorbarh}
\end{equation}
where 
\begin{equation}
\big[{\bf M}^{tt}\big]_{ij} = \delta _{ij}+ \sum _{\rho=1}^p (\lambda^\rho -1)\, v_i^\rho\, v_j^\rho 
+ \sum _{\rho=1}^{\hat p} (\hat\lambda^\rho -1)\, \hat v_i^\rho\, \hat v_j^\rho \ .
\end{equation}
The expressions for the fluctuations of the pattern components are reported in (\ref{errorbar}). Note that the cross-term $\langle \Delta T_i  \,\Delta \xi_j^\nu\rangle$ vanishes at the expected order of $\frac {\sqrt N}B$, and is actually of the order of $\frac 1{B}$ only. Using formula (\ref{changevar}) we find that the error over the fields $h_i$ is of the order of $\frac p{\sqrt \alpha}$, where $\alpha=\frac BN$.

\subsection{Optimal number of patterns}\label{secoptim}

So far we have assumed that the number of patterns, $p$, was known. In practice $p$ is often determined based on simple criteria, such as how many eigenvalues 'come out' from the spectrum of the correlation matrix (Section~\ref{rl}). Alternative approaches exist, {\em e.g.} Bayesian Information Criterion (BIC) \cite{schwarz}. In the BIC the decrease $B\Delta \Phi$  (\ref{gainphi}) in cross entropy obtained with a new pattern is added a 'cost' $N\log B$, equal to the number of new parameters times the logarithm of the number of data. As the index $\mu$ increases the selected eigenvalue $\lambda^\mu$ or $\hat \lambda^\mu$ gets closer to one; $B|\Delta \Phi|$ (\ref{gainphi}) decreases in absolute value, and, eventually, is counterbalanced by the cost term $N\log B$. The value of $\mu$ for which the two terms balance each other depends on the size of the data set: the higher $B$, the more significative are the correlations and the more patterns we need to represent the interactions. However BIC is mathematically justified when $B$ is large compared to $N$, which is not always the case in real data sets.

Hereafter, we propose a different approach based on Bayesian and geometric considerations. Based on the discussion in Section \ref{secresupatt} we expect the squared norm $b^\mu$ of the transerve fluctuations $\boldsymbol\beta^\mu$ to be non vanishing in the $B,N\to\infty$ limits. Let us call $a^\mu$ the squared projection of the $\mu^{th}$ rescaled pattern onto ${\bf v}^\mu$ (\ref{fluc78}). The same quantities, $\hat a^\nu$ and $\hat b^\nu$, can be defined for repulsive patterns. We define the marginal probability $P_M$ of the squared projections $a^\mu,\hat a^\nu$ and of the squared norms $b^\mu,\hat b^\nu$ through
\begin{eqnarray}
P_M  &=&  \int \prod _{\mu,i} \frac{d\beta^\mu_i}{\sqrt{1-m_i^2}} \prod _{\nu,i} \frac{d\hat \beta^\nu_i}{\sqrt{1-m_i^2}} \,\prod_\mu \frac{d\Omega^\mu}{\pi i \alpha N/2}\,
\nonumber \\ &\times &\prod_\nu \frac{d\hat\Omega^\nu}{\pi i \alpha N/2}\ \exp\bigg[-  \frac{\alpha}{2} \sum_\mu \Omega ^\mu \, \big( ({\boldsymbol \beta}^\mu)^2-N\, b^\mu\big) \bigg]
\nonumber\\
&\times&
\exp\bigg[-  \frac{\alpha}{2} \sum_\nu \hat \Omega ^\nu \, \big( (\hat{\boldsymbol \beta}^\nu)^2-N\,\hat b^\nu\big) \bigg]\\
&\times&P\bigg[\{T_i^0,\frac{\sqrt {N\,a^\mu} \; v_i^\mu + \sqrt N\beta^\mu _i}{\sqrt{1-m_i^2}},\frac{\sqrt {N\,\hat a^\nu} \; \hat v_i^\nu + \sqrt N\hat \beta^\nu _i}{\sqrt{1-m_i^2}}\}\bigg]
\ ,\nonumber
\end{eqnarray}
where $P$ is the posterior probability (\ref{post}), and the sums over $\mu$ and $\nu$ run from 1 to, respectively, $p$ and $\hat p$. After carrying out the integrals over the fluctuations ${\boldsymbol \beta}^\mu$ and $\hat{\boldsymbol \beta}^\nu$ we obtain
\begin{eqnarray}\label{trans}
P_M&=& \frac 1{Z_1}\int \prod_\mu  d\Omega^\mu  \prod_\nu  d\hat\Omega^\nu \\
&\times & \exp\left[ -\frac B2\, \sum _\mu \Delta\Phi_M(\Omega^\mu)  -\frac B2\, \sum _\nu
\Delta\hat \Phi_M(\hat \Omega^\nu)\right]\nonumber
\end{eqnarray}
where $Z_1$ is a normalization constant and
\begin{eqnarray}
\Delta\Phi_M (\Omega^\mu) &=&  \lambda^\mu \, a^\mu +  \Omega^\mu \, b^\mu+\log \left( 1 -a^\mu- b^\mu\right) \\&-& \frac 1{B}\log \det \big[ \Omega^\mu \, {\bf 1} - \Gamma^{(r)}\big] 
+ O\big( \frac {\log N}{N}\big)\ ,\nonumber \\
\Delta\hat \Phi_M (\hat\Omega^\nu) &=& -\hat \lambda^\nu \, \hat a^\nu + \hat \Omega^\nu \,\hat b^\nu+\log \left( 1 +\hat a^\nu+ \hat b^\nu\right) \\&-& \frac 1{B}\log \det \big[ \hat\Omega^\nu \, {\bf 1} + \Gamma^{(r)}\big] 
+ O\big( \frac {\log N}{N}\big)\ ,\nonumber
\end{eqnarray}
Here ${\bf 1}$ denotes the $N$-dimensional identity matrix. When $B$ is large the integrals in (\ref{trans}) are dominated by the contributions coming from the vicinity of the roots of 
\begin{equation}\label{calculbmu}
\frac{\partial\Delta\Phi_M}{\partial \Omega^\mu} = \frac{\partial\Delta\hat \Phi_M}{\partial\hat \Omega^\nu} = 0 \ . 
\end{equation}
Maximimization of $\Delta\Phi_M$ with respect to the $a^\mu,b^\mu$'s gives equations (\ref{map1}) and 
\begin{equation}
\Omega^\mu  = \lambda^\mu\ , \label{map3}
\end{equation}
for each $\mu=1,\ldots,p$.
We then compute the squared norm $b^\mu$ from the extremization condition (\ref{calculbmu}) and obtain
\begin{eqnarray}
b^\mu &=& \frac 1B\sum _{k=p+1}^{N-\hat p} \frac 1{\lambda^\mu-\lambda^k} \ , \label{aaabbb} \\
a^\mu&=& 1-\frac1{\lambda^\mu}-b^\mu\ . \label{amumap2}
\end{eqnarray}
Repeating the same procedure to maximize $\Delta \hat \Phi_M$ gives 
\begin{eqnarray}\label{amumap2c}
\hat {b}^\nu &=& \frac 1B\sum _{k=p+1}^{N-\hat p} \frac 1{\lambda^k-\hat \lambda^\nu}\ ,\nonumber  \\
\hat{a}^\nu&=& \frac1{\hat{\lambda}^\nu}-1-\hat{b}^\nu\ . 
\end{eqnarray}

The difference between expressions (\ref{amumap1}) and (\ref{amumap2}) for the coefficients $a^\mu$ must be emphasized. $P$ defined in (\ref{post}) is a probability density over $pN$ pattern components, once the pseudo-magnetizations have been inferred. Maximization of $P$, or, equivalently, of $\Phi$ over this large-dimensional space gives expression (\ref{amumap1}) for the projection $a^{\mu}$ of the pattern ${\boldsymbol \xi}^\mu$ onto the $\mu^{th}$ largest eigenvector of $\Gamma$, ${\bf v}^\mu$. Instead of directly maximizing $P$, we may first integrate out the orthogonal fluctuations to ${\bf v}^{\mu}$ in $P$, and obtain the marginal probability density $P_M$ for $2p$ parameters only, namely the squared projections on the eigenvectors, $a^\mu$, and the squared norms of the orthogonal fluctuations, $b^\mu$. Maximizing the marginal probability density $P_M$ or, equivalently, minimizing $\Phi_M$ shows that $b^\mu$ (\ref{amumap2}) does not vanish, and that the value of the squared projection $a^\mu$ (\ref{amumap2}) is smaller than (\ref{amumap1}). Figure \ref{fig-schema0} sketches the geometrical meaning of the coefficient $\sqrt{a^\mu}$ and the fluctuations $\boldsymbol\beta^\mu$, see (\ref{fluc78}). Small values of the angle $\theta^\mu$ are expected for reliable patterns. A similar picture can be drawn for repulsive patterns. We will see how expression (\ref{amumap2}) for the squared norm $b^\mu$ naturally arises in the context of random matrix theory.

\subsection{Maximum likelihood inference: first corrections}\label{foh}

We now look for the corrections to the lowest order expressions of the patterns and the fields (\ref{ordre0xi},\ref{ordre0T}), encoded in expressions (\ref{correxp}) and $T_i=T_i^0+T_i^1$. The first order contribution to the cross entropy, $\Phi^1$, can be seen as a perturbation to the lowest order cross entropy, $\Phi^0$, according to (\ref{expphi}). Within linear response theory this perturbation will shift the maximum likelihood estimators by 
\begin{equation}
\left( \begin{array}c {\bf T^1} \\ \{(\boldsymbol \xi^1)^{\mu}\} \\ \{\hat{(\boldsymbol \xi}^1)^{\mu}\} \end{array}\right) = - \big({\bf H}\big) ^{-1} \ \left( \begin{array}c \frac{\partial \Phi^1}{\partial\bf T} \\ \{ \frac{\partial \Phi^1}{\partial \boldsymbol \xi^{\mu}}\} \\  \{ \frac{\partial \Phi^1}{\partial \hat{\boldsymbol \xi^{\mu}}}\} \end{array}\right)
\ ,
\end{equation}
where the inverse of the Hessian matrix of $\Phi^0$, ${\bf H}$, was given in Section \ref{secerrcal}. The calculation of the gradient of $\Phi^1$ does not present any particular difficulty. The resulting corrections to the patterns are given in eqn (\ref{ordre1xi}). The expression for the shift in the pseudo-magnetization is
\begin{eqnarray}\label{ordre1h}
T^1_i &=& \sum _{\mu=1}^p (\lambda^\mu-1) \; \bigg[ C^\mu \, v_i^\mu\,  \sqrt{1-m_i^2}+ m_i\, (v_i^\mu)^2\bigg]  \\ &+& \sum _{\mu =1}^{\hat p} (\hat \lambda^\mu-1) \; \bigg[ C^{N+1-\mu} \, \hat v_i^\mu\,  \sqrt{1-m_i^2}+ m_i\, (\hat v_i^\mu)^2\bigg]\ .\nonumber
\end{eqnarray}
where $C^k$ is given in (\ref{defC3}). Notice that, if the magnetizations $m_i$ vanish, so do the dominant and first-order contributions to the pseudo-magnetizations.

\section{Reliability of the inference}\label{size}

An important issue is to determine how many configurations should be sampled in order to ensure that the inference of the patterns is accurate. To do so, we assume that the examples $\boldsymbol\sigma^b$ are drawn independently and at random from the equilibrium probability $P_{H}$ (\ref{likelihood}) of a Hopfield model, with fixed fields $\tilde{\bf h}$ and patterns $\tilde {\boldsymbol \xi}$. We call $S[\{\boldsymbol \sigma ^b\}]$ the entropy of the posterior distribution $P$ (\ref{post}) for the fields ${\bf h}$ and patterns $\boldsymbol \xi$. In the large $N$ limit, we expect this entropy to be self-averaging, that is, to depend on the set of examples only through their number $B$. We want to determine how fast $S$ decays with $B$. To do so it is instructive to first consider the simple case where the local fields are known, and only one pattern has to be inferred. This specific situation is treated in great analytical details in Section \ref{zeroh}. The general (and harder) case where both fields and patterns have to inferred is treated in Section \ref{nonzeroh}.

\subsection{Case of one unknown pattern and known fields}
\label{zeroh}

Throughout this Section, we assume that the local fields vanish, $\tilde{\bf h}=0$ and that the number of patterns to be inferred is $p=1$. The posterior entropy, 
\begin{equation}\label{defs1}
S[\{{\boldsymbol\sigma}^b\}]= -\sum _{\{\xi_i = \pm \tilde \xi\}} P[0, {\boldsymbol \xi}|\{{\boldsymbol\sigma}^b\}] \; \log
P[0, {\boldsymbol \xi}|\{{\boldsymbol\sigma}^b\}] \ ,
\end{equation}
therefore measures the uncertainty about this unique pattern given a set $B$ sampled configurations. Intuitively, the dependence of $S$ on $B$ is closely related to the physics of the Hopfield model (with pattern $\tilde{\boldsymbol\xi}$ and zero fields) used to generate the examples. If the model is in the paramagnetic phase, {\em i.e.} if the components of the pattern are weak \cite{amit85}, the examples $\boldsymbol\sigma^b$ have vanishingly small overlap (\ref{defqmu}) with the pattern. We expect that a large number $B$ (diverging with $N$) of examples is necessary to convey reliable information about the pattern. Conversely, few configurations sampled in a ferromagnetic state around a strong pattern (or its opposite) should be sufficient to reconstruct the pattern. 

We now make this scenario quantitative in various cases. An important simplication arises when the pattern is restricted to have binary components, $\tilde{\boldsymbol \xi}=\{\tilde\xi_{i}=\pm \tilde\xi\}$, with $\tilde \xi>0$. Hamiltonian (\ref{energy}) with $p=1$ pattern is invariant under the exchange of the spin configuration and the pattern: $E[\boldsymbol \sigma, 0,\boldsymbol \xi]=E[\boldsymbol \xi, 0, \boldsymbol \sigma]$. Our inference problem can thus be mapped onto a dual Hopfield model, where the normalized inferred pattern, $\boldsymbol\xi/\tilde\xi$, plays the role of the dual spin configuration and the sampled spin configurations, $\boldsymbol\sigma^b$, $b=1,\ldots , B$ correspond to the $B$ dual patterns. In particular, the posterior entropy $S$ is equal to the entropy of the dual Hopfield model at inverse temperature 
\begin{equation}\label{defbeta}
\beta =\tilde\xi^2 \ .
\end{equation}
The duality property allows us to exploit the well-understood physics of the Hopfield model \cite{amit85} to simplify the study of our inference problem.

\begin{figure}
\begin{center}
\epsfig{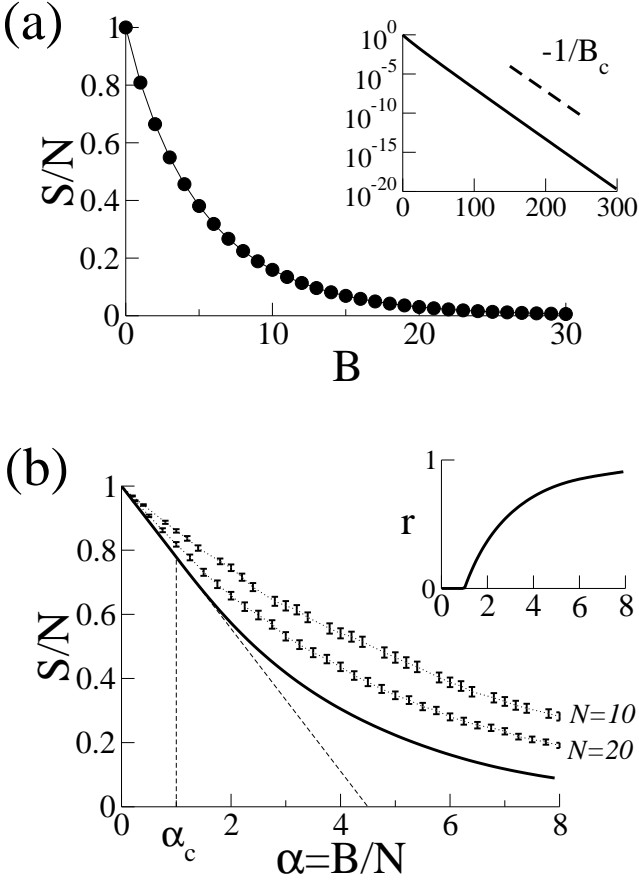}
\caption{Entropy of the posterior distribution for the patterns, $S$ (in bits and per component), as a function of the number of sampled configurations, $B$, when the local fields $h_i$ are known to vanish. {\bf (a).} Ferromagnetic regime ($\tilde\xi^2=1.1$): the entropy decays exponentially with $B$. Inset: comparison with the theoretical prediction $\exp(-B/B_{c})$ (dashed line), with $B_{c}\simeq 6.85$, in semi-log scale. {\bf (b).} Paramagnetic regime ($\tilde\xi ^2=.5$): $S$ (\ref{sentro}) is a decreasing function of $\alpha=B/N$. The entropies calculated from numerical calculations are shown for $N=10$ and $N=20$. Inset: the overlap $r$ (\ref{defq}) between the inferred and true patterns is positive when $\alpha$ exceeds $\alpha_{c}=1$ (\ref{alphac}).}
\label{fig-entropy}
\end{center}
\end{figure}

\subsubsection{Strong components}\label{strong1}

In the ferromagnetic regime ($\tilde\xi > 1$), the dual spin configuration is strongly magnetized along the dual patterns. Going back to the inference problem, we find that the overlap between the inferred pattern and a sampled configuration,
\begin{equation}\label{defm}
q^b = \sum _{\{\boldsymbol\sigma ^b\},\boldsymbol\xi} P[0, \boldsymbol \xi| \{\boldsymbol \sigma ^b\}]\prod_{b} P_H[\boldsymbol\sigma ^b,\tilde{ \boldsymbol \xi}] \; \frac 1N\sum_{i} \xi_{i} \sigma_{i}^1\ ,
\end{equation}
may take values $+q$ or $-q$, where $q$ is the positive root of $q=\tanh(q \,\tilde \xi^2)$. The sign of the overlap $q^b$ is random, depending on which one of the two states with opposite magnetizations the configuration ${\boldsymbol \sigma}^b$ in sampled in; it is equal to $+$ or $-$ with equal probabilities $\frac 12$. These statements hold if the thermodynamical limit, $N\to\infty$, is taken while $B$ is kept fixed. We find that $S$ is equal to the entropy of a single spin at inverse temperature $\beta$, interacting with $B$ other spins of magnetization $q$,
\begin{equation}\label{expresss}
S=\sum_{b=0}^B {B \choose b} \left( \frac{1+q}2 \right)^b \left( \frac{1-q}2\right)^{B-b} \; {\cal S}\big((B-2 b) q\tilde \xi^2 \big)\ ,
\end{equation}
where ${\cal S}(u)=\log ( 2 \cosh u) -u \tanh u$. Figure~\ref{fig-entropy}A shows that the entropy is almost a pure exponential: $\log S\simeq -B/B_{c}$ where the decay constant, $B_{c}=1/ \log \cosh(q \,\tilde \xi^2)$, is finite (compared to $N$). In the ferromagnetic regime few sampled configurations are sufficient to determine $\tilde{\boldsymbol\xi}$ accurately.

This result also applies to the case of a single ferromagnetic state. If the field ${\bf h}$ does not strictly vanish and explicitely breaks the reversal symmetry between the two states, all configurations are sampled from the same state, with probability $1-\exp (-O(N))$. Remarkably, expression (\ref{expresss}) for the entropy still holds. Again we find that $B=O(1)$ configurations are sufficient to infer the pattern. We will discuss in more details the inference in the ferromagnetic regime in Sections \ref{ferro1} and \ref{secferro2}.

\subsubsection{Weak components}\label{weak1}

In the paramagnetic phase ($\tilde\xi< 1$), the overlap  (\ref{defm}) between the inferred pattern and an example is typically very small, $q\sim N^{-1/2}$. No inference is possible unless the number of examples, $B$, scales linearly with $N$; we denote $\alpha=B/N$. In this regime, we expect the entropy to be self-averaging: $S[\{{\boldsymbol\sigma}^b\}]$ does not depend on the detailed composition of the data set and is a function of the value of the macroscopic parameters, {\em e.g.} the ratio $\alpha$, only. To calculate this function $S$ we use the replica method \cite{amit85,amit92}. We report below the results of the replica symmetric calculation; technical details can be found in Appendix. The order parameter is the average overlap $r$ between  the inferred and the true patterns,
\begin{equation}\label{defq}
r = \sum _{\{\boldsymbol\sigma ^b\},\boldsymbol\xi} P[0, \boldsymbol \xi| \{\boldsymbol \sigma ^b\}]\prod_{b} P_H[\boldsymbol\sigma ^b,\tilde{ \boldsymbol \xi}] \; \frac 1N\sum_{i} \xi_{i} \tilde \xi_{i}\ .
\end{equation}
which is solution of the self-consistent equation 
\begin{equation}\label{defrop}
r=\int_{-\infty}^\infty Dz \;\tanh (  z\sqrt \gamma +\gamma )\ ,
\end{equation} 
where $Dz=\frac{dz}{\sqrt{2\pi}} e^{-z^2/2}$ is the Gaussian measure, and 
\begin{equation}\label{defgop}
\gamma=\frac{\alpha \beta^2 r}{(1-\beta)(1-\beta +\beta r)}\ .
\end{equation} 
The posterior entropy is equal to
\begin{eqnarray}\label{sentro}
S&=& \int_{-\infty}^\infty  Dz\, \log 2 \cosh( z\sqrt \gamma +\gamma ) - \frac{\alpha}2\log (1-\beta +\beta r)
\nonumber \\ &-& \frac{\alpha \beta (1-\beta -r+3 \beta r)}{2 (1-\beta)(1-\beta +\beta r)} \ ,
\end{eqnarray}
and is plotted in Fig.~\ref{fig-entropy}B. To check this analytical prediction we have run extensive numerical simulations on small-size systems ($N=10,20$). The numerical procedure follows three steps: 1. evaluate the partition function $Z$ in (\ref{likelihood}) through an exact enumeration; 2. generate a data set of $B=\alpha N$ configurations $\{\sigma _i^b\}$ according to the Hopfield measure $P_H$ by rejection sampling; 3. evaluate $P_1$ in (\ref{post}) and $S$ in (\ref{defs1}) through exact enumerations. The resulting entropy, averaged over one hundred data sets, is compatible with the analytical prediction and the existence of $\frac1N$ finite-size effects. 

Inset of Fig.~\ref{fig-entropy}B shows that the overlap $r$ remains null until $\alpha$ reaches the critical value
\begin{equation}\label{alphac}
\alpha_{c} = \left( \frac 1{\tilde\xi^2}-1\right)^2 \ .
\end{equation}
Hence,  in the range $[0;\alpha_{c}]$, the posterior probability becomes more concentrated  ($S$ decreases), but not around the true pattern $\tilde{\boldsymbol \xi}$. The existence of a lagging phase before any meaningful inference is possible is similar to the 'retarded learning' phenomenon discovered in the field of unsupervised learning, where the variables to be learned are real-valued \cite{biehl94,watkin94,reimann96}. In the present case of binary spins we expect the replica symmetric assumption to break down at large $\alpha$. The entropy (\ref{sentro}) indeed becomes negative when $\alpha > \alpha_0 \simeq 42$ for the case studied in Fig.~\ref{fig-entropy}B. Nevertheless we may conjecture that the entropy decays as $S\sim \frac 1{\alpha}$ when $\alpha \to\infty$. The dual Hopfield model has random couplings $J_{ij}$, with second moment equal to $\langle J_{ij}^2\rangle -\langle J_{ij}\rangle^2=\frac{\alpha}N$. Hence $T=\frac 1{\sqrt \alpha}$ sets the temperature scale of the dual model. The low temperature scaling of the entropy of the Sherrington-Kirkpatrick (SK) model suggests that $S\propto T^2$ \cite{entroT0}; this scaling is compatible with the small--$N$ results of Fig.~\ref{fig-entropy}B. However the dual and SK models are not strictly identical when $\alpha\to\infty$: the coupling matrix $\bf J$ of the dual model is guaranteed to be semidefinite positive, while the entries of $\bf J$ are independent in the SK model. A complete calculation of the entropy valid for any (large) $\alpha$ would require a replica symmetry broken Ansatz for the order parameters \cite{dean}, and is beyond the scope of this article. 

Note that the calculations above can be extended to real patterns; $\beta$ in (\ref{defbeta}) is then replaced with $\langle \xi^2\rangle$, where the average is taken over the pattern components. The entropy is not constrained to be positive as in the binary case. The distinction between the strong- and weak-component regimes remains qualitatively unchanged, and so does the value of the critical ratio $\alpha_{c}$ (\ref{alphac}), which does not depend on the third and higher moments of $\tilde\xi_{i}$.

\subsection{General case of unknown patterns and fields}\label{nonzeroh}

In this Section, we first interpret the above results. We show that, while $B=O(1)$ configurations can be sufficient in a particular context, $B=O(N)$ data are generally necessary for the inference to be sucessful. The connection between the results of Section \ref{zeroh} and random matrix theory are emphasized. 

\subsubsection{Inference from the magnetizations}\label{ferro1}

Consider first the case where a single state exists, {\em i.e.} equations (\ref{scm}) admit a single solution $\{q^\mu \}$; the case where states coexist will be discussed in Section \ref{secferro2}. For large $N$, the average value of spin $i$ with the measure $P_H$ (\ref{likelihood}) is 
\begin{equation}\label{localmi}
m_i = \tanh \big( h_i + \sum _\mu q^\mu\, \xi_i^\mu \big) \ .
\end{equation}
As the error on the estimate of $m_i$ decreases as $\sim \sqrt{\frac{1-m_i^2}B}$ with $B$, $O(1)$ configurations are sufficient to sample the magnetizations accurately. Few sampled configurations therefore give access to the knowledge of a linear combination of the field vector and pattern vectors with non zero-overlaps $q^\mu$. This linear combination is simply $T^0_i$, and equation (\ref{localmi}) coincides with (\ref{ordre0T}). 

When the fields $h_i$ are known and the model consists of a single strong pattern ($p=1$) the pattern components $\xi^1_i$ can be readily calculated from the magnetizations (\ref{localmi}) through 
\begin{equation}
\xi_i^1 = \frac 1q\, \tanh^{-1} m_i \quad \hbox{\rm where} \quad
q^{2} = \frac 1N \sum _j m_j  \tanh^{-1} m_j \ .
\end{equation}
This particular case was encountered at the end of Section \ref{strong1}, when the fields $h_i$ are sent to zero after having broken the reversal symmetry of the system to avoid state coexistence. In the generic situation of unknown fields and patterns, knowledge of the magnetizations does not suffice to determine the field and the patterns, and must be supplemented with the information coming from the correlation matrix $\Gamma_{ij}$.

\subsubsection{Inference from the correlations: relationship with random matrix theory}\label{rl}

What is the order of magnitude of $\Gamma_{ij}$? We first consider the ideal case of perfect sampling ($B\to\infty$  while $N$ is large but finite). As a result of the presence of the patterns in the energy (\ref{energy}) the spins are correlated. The entries of the correlation matrix are, for large $N$ \footnote{Formula (\ref{cc}) can be found by inverting identity (\ref{jtap}), with $J_{ij}=\frac 1N \xi_i\xi_j$.},
\begin{equation}\label{cc}
\Gamma_{ij} =\delta_{ij} + \frac 1N \frac{\xi_i \xi_j \sqrt{(1-m_i^2)(1-m_j^2)}}{1- \frac 1N \sum _{k}  \xi_k^2 \, (1-m_k)^2 }
\end{equation}
where we have considered the case of a single pattern ($p=1,\hat p=0$) to lighten notations. Though the pattern affects each correlation $\Gamma_{ij}$ by $O(\frac 1N)$ only, these small contributions add up to boost the largest eigenvalue from one (in the absence of pattern) to
\begin{equation}\label{valuel}
L = \frac 1{1- \frac 1N \sum _{k} \xi_k^2 \, (1-m_k)^2 } \ .
\end{equation}
The eigenvector attached to $L$ has components $v_i \propto \xi_i\sqrt{1-m_i^2}$ and ML inference perfectly recovers the pattern.

In the presence of sampling noise (finite $B$), each correlation (\ref{cc}) is corrupted by a stochastic term of the order of $x=\frac 1{\sqrt B}$. This stochastic term will, in turn, produce an overall contribution of the order of $x\sqrt N=\frac 1{\sqrt\alpha}$ to the largest eigenvalue. Intuitively, whether $\alpha$ is large or small compared to $L^{-2}$ should tell us how hard or easy it is to extract the pattern $\boldsymbol\xi$ from $\Gamma$. Several studies in the physics \cite{hoyle03,hoyle10} and in the mathematics \cite{baik05} literatures have indeed found that an abrupt phase transition takes place at the critical ratio 
\begin{equation}\label{ac}
\alpha_c = \frac 1{(L-1)^2} \ .
\end{equation}
It is a simple check that $\alpha_{c}$ coincides with the ratio (\ref{alphac}) for the retarded learning transition calculated in Sections \ref{weak1}.

In the strong noise regime ($\alpha < \alpha_c$) the largest eigenvector ${\bf v}^1$ of $\Gamma$ is uncorrelated with (orthogonal to) the pattern $\boldsymbol \xi$, and the spectrum of $\Gamma$ is identical to the one of the sample correlation matrix of independent spins, whose density of eigenvalues is given by the Marcenko-Pastur (MP) law,
\begin{equation}\label{MPd}
\rho_{MP} (\lambda ') =v(1-\alpha)\, \delta (\lambda ') + \frac{\alpha}{2\pi \lambda '} \; \sqrt{v \big((\lambda_+-\lambda ')(\lambda '-\lambda_-)\big)}
\end{equation}
with $v(u)=\max (u,0)$ \cite{johnstone06}. The edges of the continuous component of the MP spectrum are given by
\begin{equation}\label{valueMP}
\lambda _\pm = \left (1 -\frac 1{\sqrt \alpha}\right)^2 \ .
\end{equation}
The largest eigenvalue of $\Gamma$, $\lambda_{+}$, is not related to the value of $L$.

In the weak noise regime ($\alpha > \alpha_{c})$ the largest eigenvalue of $\Gamma$ is \cite{baik05}
\begin{equation}\label{valuel1}
\lambda^1 = L \; \left( 1 + \frac 1{\alpha\, (L-1)}\right) \ .
\end{equation}
It exceeds $L$ for any finite $\alpha$, and converges to $L$ when $\alpha\to\infty$. The rest of the spectrum is described by the MP density (\ref{MPd}). Expression (\ref{aaabbb}) for the squared norm $b^1$ of the orthogonal fluctuations
leads to the analytical formula
\begin{equation}\label{valueb}
b^1 =\frac 1{\alpha} \int _{\lambda_-}^{\lambda_+} d\lambda '\; \frac{\rho_{MP} (\lambda')} {\lambda^1-\lambda'}= \frac{\lambda^1 -L}{\lambda^1} \ ,
\end{equation}
where we have used the analytical expression of the Stieltjes transform of $\rho_{MP}$ \cite{book}. Using (\ref{amumap2}) we deduce the value of the squared projection of the inferred rescaled pattern $({\boldsymbol \xi}^1)'$ onto ${\bf v}^1$,
\begin{equation}\label{valuea}
a^1 = \frac{L-1}{\lambda^1}  \ .
\end{equation}
Identities (\ref{valueb}) and (\ref{valuea}) are graphically interpreted in Fig.~\ref{fig-schema0}: $b^1$ is the squared norm of the orthogonal fluctuations $\boldsymbol\beta$, while $a^1$ is the squared projection of the rescaled pattern $\boldsymbol\xi$ onto ${\bf v}^1$. 

\begin{figure}{b}
\begin{center}
\epsfig{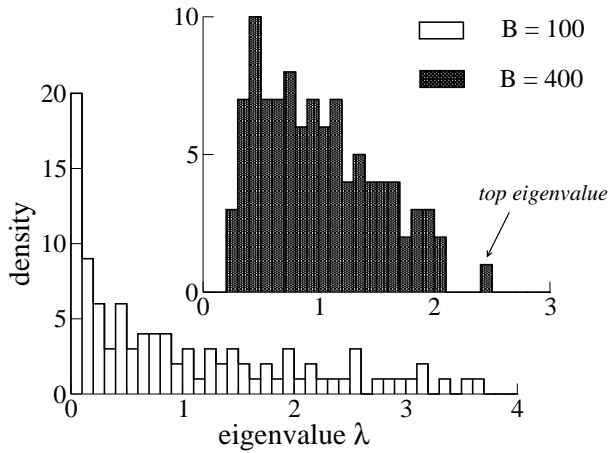}
\caption{Spectrum of the correlation matrix for a Hopfield model with $p=1$ pattern, $N=100$ spins, and for $B=100$ (bottom) and 400 (top) randomly sampled configurations at equilibrium. The bulk parts of the spectra coincide with the Marcenko-Pastur law for random correlation matrices. When $B$ is large the top eigenvalue clearly comes out from the noisy bulk and the corresponding eigenvector approximately corresponds to the pattern. The pattern components are i.i.d. Gaussian variables, of zero mean and variance $\xi^2=.5$; local fields $h_i$ have zero values.}
\label{fig-spectre}
\end{center}
\end{figure}

The above discussion is illustrated on the simple case of a Hopfield model with $p=1,\hat p=0$ patterns in Fig.~\ref{fig-spectre}, see caption for the description of the model. Using formula (\ref{valuel}) we compute the largest eigenvalue of the correlation matrix for perfect sampling, $L=2$. Figure~\ref{fig-spectre} shows that a large eigenvalue clearly pulls out from the bulk spectrum for the ratio $\alpha=4$ (top spectrum), larger than the critical ratio $\alpha_{c}=1$ according to (\ref{ac}) (bottom). For $\alpha=4$, the infinite--$N$ predicted values for the largest eigenvalue, $\lambda_{1}=2.5$  (\ref{valuel1}), and for the edges of the MP spectrum, $\lambda_{-}=.25,\lambda_{+}=2.25$ (\ref{valueMP}), are in good agreement with the numerical results for $N=100$.

Formulae (\ref{valueb}) and (\ref{valuea}) hold for each pattern $\mu$ when $p\ge 2$ patterns are present, provided that $p$ remains finite when $N\to\infty$. The case of $p=2$ patterns, where one pattern is strong and has overlap $q>0$ (\ref{defm}) with the sampled configurations, and the second pattern has weak components, is of particular interest. Again, we assume that the fields vanish. Repeating the calculation of Section \ref{weak1} and Appendix A we find that the entropy $S/N$ quickly decreases with $B$ from 2 bits down to 1 for $B=O(1)$. When $B\propto N$, the entropy decreases from 1 down to 0; the expression of $S$ coincides with (\ref{sentro}) where $\beta$ is replaced with $\beta (1-q^2)$. Hence we have a two-step behaviour: the strong pattern is determined with $O(1)$ examples, the weak pattern requires $O(N)$ sampled configurations. Learning of the weak pattern is possible if
\begin{equation}
\alpha \ge \left( \frac 1{\tilde\xi^2 (1-q^2)}-1\right)^2 \ ,
\end{equation}
according to (\ref{alphac}). The two-step behaviour agrees with the discussion of Section \ref{ferro1}.

\subsubsection{Coexistence of ferromagnetic states}\label{secferro2}

Consider now the case of the coexistence of two ferromagnetic states exposed in Section \ref{seccoex}. Data are generated from a Hopfield model, with zero fields and one strong pattern $\boldsymbol\xi$, as in Fig.~\ref{fig-compar-1patt-bis}. In the up-state the spins are magnetized with $m_i^+=\tanh(q\,\xi_i)$. In the down-state the local magnetization is $m_i^-=-m_i^+$. On the overall the local magnetization is $m_i=\frac 12\, m_i^+ + \frac 12\, m_i^-=0$, up to $O(\frac 1{\sqrt B})$ fluctuations. The discrepancy between the Gibbs magnetizations, $m_i=0$, and the state magnetizations, $m_i^\pm$, results in a $O(1)$ contribution $m_i^+ m_j^+ (=m_i^-m_j^-)$ to the correlation matrix entry $\Gamma_{ij}$, dominating the $O(\frac 1N)$ contributions due to the interactions between spins. The largest eigenvalue of $\Gamma$,
\begin{equation}
\lambda ^1 = \sum _i (m_i^+)^2\ ,
\end{equation}
is of the order of $N$; the corresponding eigenvector is ${\bf v}^1=(m_1^+,m_2^+,\ldots, m_N^+)/\sqrt{\lambda^1}$. Informally speaking, the information about the state magnetizations is not conveyed by the Gibbs magnetizations (as in Section \ref{ferro1}) but by the correlation matrix \cite{sinova}. According to formula (\ref{ordre0T}) the pseudo-magnetization $T_i$ vanishes; hence we correctly infer that the fields $h_i$ have zero values. Using formula (\ref{ordre0xi}) we obtain
\begin{equation}
(\xi^0)_i \simeq \sqrt{\frac N{\lambda^1}}\; m_i^+ \ .
\end{equation}
Therefore, the inferred pattern component is not equal to the true pattern component, but is proportional to its hyperbolic tangent. This non linear transform is clearly seen in Fig.~\ref{fig-compar-1patt-bis}. The discrepancy between the true and inferred components is a nice illustration of the claimed scaling for the higher order corrections in (\ref{expr}) (recall that the eigenvalues of $A^{-1}$ are the $p$ largest eigenvalues of $\Gamma$). In the presence of coexistent states, while $\xi^2$ is small compared to $N$, $\lambda^1$ is of the order of $N$, making the ratio $\frac{\lambda^1 {\boldsymbol\xi}^2}N$ of the order of unity. Corrections are required and shown to improve the quality of the inferred pattern in Fig.~\ref{fig-compar-patt1-corr}. 

\section{Conclusion}\label{conc}

In this paper we have studied how to infer a small-rank interaction matrix between $N$ binary variables given the average values and pairwise correlations of those variables.  We have seen that the generalized Hopfield model, where the interactions are encoded into a set of attractive and repulsive patterns $\boldsymbol \xi$, is a natural framework for Maximum Likelihood (ML) inference. Using techniques from the statistical physics of disordered systems, we have presented a systematic expansion of the log-likelihood in powers of $\lambda \frac {{\boldsymbol\xi}^2}N$, where $\lambda$ is the largest eigenvalue of the correlation matrix $\Gamma$ (\ref{defgamma}). We have then calculated the ML estimators for the patterns and the fields to the lowest and first order in this expansion in a variety of physical regimes. The lowest order is a simple extension of Principal Component Analysis, where not only the largest but also the smallest eigenmodes build in the interactions. First order corrections involve non-linear combinations of the eigenvalues and eigenvectors of $\Gamma$. We have validated our ML expressions for the patterns on synthetic data generated by Hopfield models with known patterns and fields, and by Ising models with sparse interactions. We have also presented a simple geometrical criterion for deciding the number of patterns. Those results have been discussed and compared to previous studies in the unsupervised learning and random matrix literatures. 

The quality of the inference strongly depends on the number of sampled configurations, $B$. The sampling error on each magnetization, $m_i$, and pairwise correlation, $c_{ij}$, is of the order of $B^{-1/2}$. Elementary insights from random matrix theory suggest that the resulting errors on the eigenvectors of the matrix $\Gamma$ are $\sqrt N$ times larger. The error on the inferred patterns, $\epsilon$, picks up a contribution $\sim \left(\frac NB\right)^{1/2}$ due to finite sampling, as found in Section \ref{secsam}. This scaling has several important consequences. First, inference is retarded: no information about the true couplings can be obtained unless the ratio $\frac BN$ exceeds a critical value (Sections \ref{weak1} and \ref{rl}). Secondly, for larger $B$, $\epsilon$ decreases as $B^{-1/2}$, which is confirmed by the simulations presented in Fig.~\ref{fig-compar-pca-corr}, and then saturates to the intrinsic error resulting from our approximate expressions for the patterns. The intrinsic error depends on the order in the expansion used for the calculation of the cross-entropy in Section \ref{secinf}. Note that other inference methods, looking for the local structure of the interaction network \cite{wain,noi}, may unveil strong couplings $J=O(1)$ from a much smaller number of sampled configurations, $B=O(\log N)$, and do not suffer from the retarded learning transition. 

Our study could be extended in several directions. It would be particularly interesting to consider the case of spins taking $Q>2$ values (Potts model), {\em e.g.} for applications to the study of coevolution between residues in protein sequences \cite{loc99,riv09,weigt}. Mean-field inference methods provide a simple and efficient way to get interactions from correlations \cite{weigt10}. Knowing how MF interactions are modified when some eigenmodes are rejected (using the criterion of Section \ref{secresupatt}) or first-order corrections are taken into account would be of interest. However the linear increase in the number of possible symbols with $Q$ ($=20$ for amino-acids) may make the effective size of the problem, $N\times Q$, larger than the number of configurations, $B$, in practical applications. A large number of vanishing eigenvalues is expected in those cases, and extracting repulsive patterns may become a difficult task.

Appropriate priors $P_0$ could also be used to force many pattern components to identically vanish, instead of acquiring small values as in Section \ref{secregu}. This can be particularly useful when the true patterns are known to be highly sparse and few data are available. Inspired by the so-called Lasso regression method \cite{tish}, a natural prior is 
\begin{equation}\label{regu2}
P_0 \propto\exp \left[- \gamma\sum_{i=1}^N  \sqrt{1-m_i^2} \left(\sum_{\mu=1}^p |\xi_i^\mu|+ \sum_{\mu=1}^{\hat p} |\hat \xi_i^\mu| \right)\right]\ .
\end{equation}
Contrary to the case of the quadratic penalty (\ref{regu1}) the most likely values for the patterns cannot be expressed by means of simple analytical formulae. However, they could be efficiently obtained using convex optimization algorithms minimizing the sum of the cross entropy and of the penalty term (\ref{regu2}). 
 
Last of all, we have considered in this work that the configurations were sampled at equilibrium. In practice, when more than one state exist, the equilibration time may be prohibitive and a reasonable assumption would be to sample from one state only. To what extent ergodicity breaking in the sampling affects the quality of inference is an interesting question.

\vskip .3cm
{\bf Acknowledgments:} We thank S. Leibler for numerous discussions. V.S. thanks the Simons Center for Systems Biology for its hospitality. This work was partially funded by the ANR contract 06-JC-JC-051.

\appendix

\section{Replica calculation of the entropy $S$ for weak patterns}

When the pattern has binary components $\tilde \xi_i=\pm \tilde \xi$ we make the change of variables $\sigma_i'=\xi_i \sigma_i$ to rewrite the partition function (\ref{pf}) of the Hopfield model through
\begin{equation}
Z =  \sum_{\{\sigma'\}} \exp\left[\frac{\beta}{N}\sum_{i<j} \sigma_i' \sigma_j' +\frac {\beta}{2N}\right] \ ,
\label{eq_mattis_cw}
\end{equation}
where the inverse temperature $\beta$ is defined in (\ref{defbeta}). The partition function is thus independent of the pattern direction, which makes the calculation considerably simpler. The posterior entropy (\ref{defs1}) can be written as
\begin{equation}
S[\{\boldsymbol\sigma ^b\}] = \left( 1 - \beta \;\frac{\partial}{\partial \beta}\right) \log \tilde N [\{\boldsymbol\sigma ^b\},\beta ] \, .
\label{eq_entro11}
\end{equation}
where
\begin{equation}
\tilde{N}[\{\boldsymbol\sigma ^b\},\beta] = \sum_{\{\boldsymbol\xi\}}\exp \left(\frac{\beta}{N} \sum_{b=0}^B \sum_{i<j} \xi_i \xi_j \sigma_i^b  \sigma_j^b \right)\, ,
\label{eq_ntil}
\end{equation}
Thus, we are left with the calculation of $\tilde{N}[\{\boldsymbol\sigma ^b\}]$.  The expression for $\tilde{N}$ is formally identical to the partition function of a dual Hopfield model where the $B$ measured configurations $\boldsymbol\sigma ^b$ play the role of the dual patterns and $\boldsymbol\xi$ plays the role of the dual spin variables. The posterior entropy $S$ is simply the entropy of this dual Hopfield model. 

Equation (\ref{eq_entro11}) gives the entropy of the system for a particular set of measures $\{\boldsymbol\sigma^b\}$. It is natural to expect the entropy to be reproducible across different sets of measurements. In this context, we are interested in evaluating the average of the entropy with respect to all possible measurements. Assuming that the configurations  $\{\boldsymbol\sigma^b\}$ are sampled from the equilibrium measure of a Hopfield model with one pattern $\tilde{\boldsymbol\xi}$, we write the average entropy as
\begin{equation}
S= \left( 1 - \beta \;\frac{\partial}{\partial \beta}\right) \left. \langle \log \tilde{N}\rangle (\tilde \beta ,\beta ) \right|_{\tilde \beta=\beta }\, .
\label{eq_entro11bis}
\end{equation}
where
\begin{eqnarray}\label{a5}
\langle \log \tilde N\rangle (\tilde \beta,\beta)&=& \frac{1}{Z^B} \sum_{\{\boldsymbol\sigma ^b\}}
\exp\left(\frac{\tilde{\beta}}{N}  \sum_{b=0}^B \sum_{i<j} \tilde{\xi_i} \tilde{\xi_j} \sigma_i^b  \sigma_j^b\right) \nonumber\\
&\times &\log \tilde{N}[\{\boldsymbol\sigma^b\},\beta] \, ,
\end{eqnarray}
where we have introduced a new variable $\tilde{\beta}$ since we should not take the derivative only with respect to $\beta$ in (\ref{eq_entro11bis}).

To calculate the average value of the logarithm of $\tilde N$ in (\ref{a5}) we use the replica trick \cite{amit85} and estimate the $n^{th}$ moment of $\tilde N$,
\begin{eqnarray}
\langle\tilde{N}^n\rangle &=& e^{-\beta B n/2} \sum_{\{\boldsymbol\xi ^\rho\} , \tilde{\boldsymbol\xi},\{\boldsymbol\sigma ^b\}} \int \prod_{b=1}^B \prod_{\rho=1}^n \frac{d m_b^\rho}{\sqrt{2 \pi}}  \nonumber \\
&\times & \exp \left[- \frac{\beta N}{2} \sum_{b, \rho} \left(m^\rho_b\right)^2 
+  \beta\sum_{b,\rho,i} m_b^\rho  \xi_i^\rho \sigma_i^b  \right.
\nonumber\\
&&+ \left. 
 \frac{\tilde{\beta}}{N} \sum_b \sum_{i<j} \sigma_i^b \sigma_j^b \tilde{\xi}_i \tilde{\xi}_j\right] \ .
\end{eqnarray}
We introduce auxiliary Gaussian variables, denoted by $\tilde m_{b}$, to linearize the quadratic term in the spins $\sigma _{i}^b$. We obtain, after summation over the spins,
\begin{eqnarray}
\langle\tilde{N}^n\rangle &=& e^{-\beta B n/2} \sum_{\{\boldsymbol \xi ^\rho\}, \tilde{\boldsymbol \xi}} \int \prod_{b, \rho}
\frac{d m_b^\rho}{\sqrt{2 \pi}} \prod _{b} \frac{d \tilde{m}_b}{\sqrt{2 \pi} }\\
&\times &  \exp \left[ - \frac{\beta N}{2} \sum_{b, \rho} \left(m^\rho_b\right)^2 
 - \frac{\beta N}{2} \sum_{b} \left(\tilde{m}_b \right)^2 \right.\nonumber\\
&+&\left. \sum_{i,b} \ln 2 \cosh \left(\beta \sum_\rho m^\rho_b \xi_i^\rho + 
\tilde{\beta} \,\tilde{m}_b \,\tilde{\xi}_i\right) \right] \, .\nonumber 
\end{eqnarray}
In the paramagnetic phase we expect the variables $m_{b}^\rho$ and $\tilde m_{b}$ to be of the order of $\frac  1{\sqrt N}$. Expanding the hyperbolic cosine to the second order in those variables and carrying out the resulting Gaussian integral we obtain
\begin{equation}
\langle \tilde{N}^n \rangle \simeq e^{-\beta B n/2} \sum_{\{\boldsymbol \xi ^\rho\}, \tilde{\boldsymbol \xi}}
 \left[ \det M \right]^{-B/2}\, .
\end{equation}
Here, $M$ is the $(n+1)\times (n+1)$ matrix with elements
\begin{equation}
M_{\rho\sigma} = \left\{
\begin{array}{c c c}
1- \beta & \hbox{\rm if} & \rho=\sigma \le p\ , \\
1- \tilde{\beta} & \hbox{\rm if} & \rho=\sigma=p+1\ , \\
- \sqrt{\beta \tilde{\beta}}\, t_\sigma & \hbox{\rm if} & \rho=p+1,\ \sigma\le p\ , \\
- \sqrt{\beta \tilde{\beta}}\, t_\rho & \hbox{\rm if} & \rho\le p,\ \sigma = p+1\ , \\
- \beta \, r_{\rho\sigma}  & \hbox{\rm if} & \rho \le p,\ \sigma \le p\ .
\end{array}
\right. 
\end{equation}
with the overlaps defined through $r_{\rho \sigma} = \frac{1}{N}\sum_i \xi_i^\rho \xi_i^\sigma$ and $t_\rho = \frac{1}{N}\sum_i \xi_i^\rho \tilde{\xi}_i$. We now enforce the definitions of the overlaps using conjugated Lagrange multipliers, $\hat{r}_{\rho \sigma}$ and $\hat{t}_\rho$, and obtain
\begin{eqnarray}
\langle {\tilde{N}^n} \rangle &=& \int \prod_{\rho < \sigma} \frac{dr_{\rho \sigma}\, 
d\hat{r}_{\rho \sigma}}{2\pi}\, \prod_\rho \frac{dt_\rho \, d\hat{t}_\rho}{2\pi}\  \Xi^N \, ,
\end{eqnarray}
where $\Xi$ is given by
\begin{eqnarray}
\Xi&=& \sum_{\{\xi ^\rho, \tilde{\xi}\}} \exp \left[
- \frac{\alpha}{2} \log \det M 
-\sum_{\rho<\sigma} \hat{r}_{\rho \sigma} r_{\rho \sigma}
-\frac{\alpha \beta n}{2}\right. \nonumber \\
&-& \left. \sum_\rho \hat{t}_\rho t_\rho
+  \sum_{\rho < \sigma}  \hat{r}_{\rho \sigma}  \xi^\rho \xi^\sigma + \sum_\rho \hat{t}_\rho \,
\tilde{\xi} \xi^\rho 
\right] \, .
\end{eqnarray}
We look for a replica-symmetric saddle point of $\Xi$: $r_{\rho \sigma} = r$, $t_\rho=t$, $\hat{r}_{\rho \sigma} = \hat{r}$ and $\hat{t}_\rho=\hat{t}$. We obtain, after some elementary algebra,

\begin{eqnarray}
\Xi &=& \int_{-\infty}^\infty Dz
\exp \left\{ - \frac{\alpha}{2} \log \det M 
-\frac{n (n-1)}{2}  \, \hat{r}\, r - {n}\, \hat{t}\, t\right. \nonumber \\
&+& \left. n\log\left[2\cosh\left(\hat{t}
+z\sqrt{\hat{r}}   \right)\right]- \frac{\alpha \beta n}{2}\right\} \, .
\end{eqnarray}
where $Dz=dz \,e^{-z^2/2}/\sqrt{2\pi}$ is the Gaussian measure and 
\begin{eqnarray}
\det M &=& \left(1-\beta + \beta r\right)^{n-1} \; \big[(1-\tilde{\beta})(1-\beta)
 \nonumber\\
&-& (n-1) (1-\tilde{\beta}) \beta r - n \beta \tilde{\beta}\, t^2\big] \ .
\end{eqnarray}
We now send $n$ to zero. The saddle-point equations show that $t=r$; this result was expected from the fact that, if $\tilde\beta=\beta$, the true pattern $\tilde \xi$ plays the role of an extra replicated pattern $\xi$. In addition, $\hat t=\hat r \equiv \gamma$, where $\gamma$ is defined in (\ref{defgop}). The self-consistent equations for $r$ and the entropy $S$ are given by, respectively eqns (\ref{defrop}) and (\ref{sentro}).

\end{document}